\documentclass[11pt,letterpaper]{article}
\pdfoutput=1
\usepackage{soul}
\usepackage{jheppub}
\usepackage{amsfonts,subcaption}
\usepackage{amsmath}
\allowdisplaybreaks[4]
\usepackage{amssymb}
\usepackage{euscript}
\usepackage{color}
\usepackage{tikz}
\usepackage{pgfplots}
\usetikzlibrary{intersections, pgfplots.fillbetween}

\newcommand{\bea}{\begin{eqnarray}}
\newcommand{\eea}{\end{eqnarray}}
\newcommand{\ba}{\begin{eqnarray}}
\usepackage{braket}
\newcommand{\ea}{\end{eqnarray}}

\newcommand{\beq}{\begin{equation}}
\newcommand{\eeq}{\end{equation} }
\newcommand{\beqa}{\begin{eqnarray}}
\DeclareMathOperator{\Tr}{Tr}
\newcommand{\eeqa}{\end{eqnarray}}
\newcommand{\beqar}{\begin{eqnarray*}}
\newcommand{\eeqar}{\end{eqnarray*}}

\newcommand{\be}{\begin{equation}}
\newcommand{\ee}{\end{equation}}

\newcommand{\hide}[1]{}

\def\ket#1{\mathinner{|{#1}\rangle}}

\title{Traversability of Multi-Boundary Wormholes}

\author[a]{Abdulrahim Al Balushi,}
\author[b]{Zhencheng Wang}
\author[b]{and Donald Marolf}
\emailAdd{a2albalu@uwaterloo.ca}
\emailAdd{zhencheng@ucsb.edu}
\emailAdd{marolf@ucsb.edu}
\affiliation[a]{Department of Physics and Astronomy, University of Waterloo, Waterloo, Ontario, N2L 3G1, Canada}
\affiliation[b]{Department of Physics, University of California, Santa Barbara, CA 93106, USA}

\date{\today}

\abstract{We generalize the Gao-Jafferis-Wall construction of traversable two-sided wormholes to multi-boundary wormholes. In our construction, we take the background spacetime to be multi-boundary black holes in AdS$_3$.  We work in the hot limit where the dual CFT state in certain regions locally resembles the thermofield double state. Furthermore, in these regions, the hot limit makes the causal shadow exponentially small.  Based on these two features of the hot limit, and with the three-boundary wormhole as our main example,  we show that traversability between any two asymptotic regions in a multi-boundary wormhole can be triggered using a double-trace deformation.  In particular, the two boundary regions need not have the same temperature and angular momentum.  We discuss the non-trivial angular dependence of traversability in our construction, as well as the effect of the causal shadow region.}

\begin{document}
\maketitle
\flushbottom


\section{Introduction}
\label{sec:Intro}
Wormholes have long been of interest since the time of Einstein and Rosen \cite{Einstein:1935tc}. Although Einstein-Rosen bridges connect different asymptotic regions of spacetime, topological censorship \cite{PhysRevLett.71.1486, Galloway:1999bp} forbids their traversability when only classical matter fields are present.  The same is of course true of their multi-boundary wormhole generalizations.  However in some cases, quantum matter fields can cause violations of the averaged null energy condition (ANEC).  In such cases the arguments of  \cite{PhysRevLett.71.1486, Galloway:1999bp} cannot be applied, so that such ANEC violations might make the wormholes traversable.  We remind the reader that the ANEC is satisfied when the integral of stress tensor along any complete null geodesic is non-negative,
\begin{equation}
\int_\gamma T_{ab}k^a k^b \geq 0.
\end{equation}

In recent years, there have been many approaches to constructing traversable wormholes from ANEC violations, see \cite{Gao:2016bin, Caceres:2018ehr, Fu:2018oaq,Fu:2019vco, Maldacena:2018gjk, Maldacena:2017axo, Maldacena:2018lmt, Horowitz:2019hgb}. In particular, in the seminal paper by Gao, Jafferis and Wall \citep{Gao:2016bin}, the authors construct a traversable wormhole using a two-sided BTZ black hole as the background, where the dual CFT state is the thermofield double (TFD) state. With an appropriate sign of coupling, a double-trace deformation that directly couples the two boundary CFTs can cause the violation of the ANEC. Adding the coupling shifts the horizons so as to allow certain causal geodesics to travel from one asymptotic boundary to the other. In \citep{Caceres:2018ehr}, this construction was generalized to rotating BTZ black holes. It is also interesting to recall that the transmission of such signals was interpreted in \cite{Maldacena:2017axo} from the dual field theory perspective as being due to enacting a quantum teleportation protocol between entangled quantum systems. This connection with quantum information has been of great interest (see e.g. \cite{Susskind:2017nto,vanBreukelen:2017dul,Yoshida:2018vly,Bak:2018txn,Freivogel:2019whb,Brown:2019hmk}) as a concrete realization of the ER=EPR idea \cite{Maldacena:2013xja}.

In the current paper, we generalize this construction to any pair of asymptotic regions in certain (non-rotating or rotating) multi-boundary black holes\footnote{Note that, while there is some freedom in the use of such terms, our choice is to use ``multi-boundary black holes" when the context refers to the background spacetime, and use ``multi-boundary wormholes" when the context refers to traversable wormholes in particular.}  in AdS$_3$. For a general multi-boundary black hole, a finite-sized causal shadow separates the horizons of different asymptotic regions, making the wormhole hard to traverse. In our construction, we focus on the hot limit considered in \cite{Marolf:2015vma}, where the temperatures related to all horizons are large.  In that limit, for any two horizons, there exists a region where the causal shadow between them is exponentially small. A double-trace deformation can then easily render the wormhole traversable. As we will see, the hot limit will also give us convenience in doing the calculations, which otherwise would be difficult to perform.

Our construction has several interesting features that differ from those of \citep{Gao:2016bin} and \citep{Caceres:2018ehr}. The first is that the pair of boundaries in our traversable wormhole construction is quite general, and the associated horizons can have different temperatures and angular momenta. Furthermore, our spacetimes have non-trivial angular dependence, and this can be seen in features related to traversability.  In particular, signals from a given asymptotic region will be able to reach a second asymptotic region only when fired from appropriate regions of the first boundary.  Signals launched from other parts of the first boundary may instead traverse to a third asymptotic region, or they may be become stuck behind an event horizon.  It is a general feature of our construction that some such event horizon will remain even though our wormholes are traversable.  Again, this is associated with the lack of rotational symmetry in our spacetimes.

In section \ref{sec:AdS3}, we review the construction of multi-boundary wormholes in AdS$_3$ and their important properties that will be useful in later sections. The geometry of these wormholes in the hot limit is also discussed, as well as the entanglement structure of the dual CFT state. A general review of the Gao-Jafferis-Wall construction is then given in section \ref{sec:TravReview}, where we emphasize a rather general form of the coupling between boundaries that can induce traversability. Using these two ingredients, we proceed to construct the multi-boundary traversable wormhole in section \ref{sec:MultiTrav}. We summarize our findings and discuss their implications and connections with recent work in the literature in section \ref{sec:discussion}. A number of technical details and supporting calculations are left to the appendices.

\section{Multi-boundary black holes in AdS$_3$}
\label{sec:AdS3}
In this section, we will first review how to construct multi-boundary black holes by quotienting empty AdS$_3$ with isometries, following an algebraic approach \cite{Aminneborg:1997pz,Brill:1995jv, Brill:1998pr, Maxfield:2014kra, Aminneborg:1998si,Krasnov:2001va}\footnote{For construction of these geometries using explicit forms of the Killing vectors, see \cite{Caceres:2019giy}.}. Then we discuss fixed points of those isometries, (renormalized) geodesic distances in different conformal frames, and how they behave in the hot limit. Those results will be useful in our construction of multi-boundary traversable wormholes. Finally, we briefly describe the CFT states that are dual to these geometries.

\subsection{Quotients of AdS$_3$ space}\label{sec2.1}
In three-dimensional Einstein gravity, the Ricci tensor completely specifies the Riemann tensor. The consequence of this is that all gravity solutions are locally isometric to AdS$_3$, which is the Lorentzian, maximally-symmetric spacetime with constant negative curvature and isometry group $SO(2,2)\simeq SL(2,\mathbb{R})\times SL(2,\mathbb{R})$. Besides pure AdS$_3$, other solutions to the equations of motion are locally AdS$_3$ but differ globally from it and can be obtained by quotienting AdS$_3$ by a discrete subgroup $\Gamma$ of $SO(2,2)$. Throughout the paper, we take the AdS radius $L_{AdS}=1$. The spacetime AdS$_3$ can be defined as the submanifold of
\be\label{pembedd}
\mathbb{R}^{2,2}=\left\{p=\begin{pmatrix}
U+X& -V+Y\\
V+Y& U-X
\end{pmatrix}\right\},\quad ds^2=-\det(dp)\equiv\bar{\eta}_{ab}d\bar{x}^ad\bar{x}^b,
\ee
given by the hyperboloid $\det(p)=1$\footnote{$dp$ is the matrix defined by taking the differential of every element of the matrix $p$.}, where we defined the 4-vector $\bar{x}^a=\left(U,V,X,Y\right)$ and metric $\bar{\eta}_{ab}=\text{diag}\left(-1,-1,1,1\right)$. In global coordinates, this hyperboloid is parametrized by the intrinsic coordinates $(t,r,\phi)$ defined by
\be\label{globalAdS3}
X=r\cos\phi,\quad Y=r\sin\phi,\quad U=\sqrt{1+r^2}\cos t,\quad V=\sqrt{1+r^2}\sin t
\ee
which gives the induced metric
\be
ds^2=-(1+r^2)dt^2+\frac{dr^2}{1+r^2}+r^2d\phi^2
\ee
where $t\sim t+2\pi$\footnote{Usually the universal cover of $t$ is taken by unwrapping it, but as we will see, it is not necessary here since the wormhole constructions will automatically remove closed timelike curves.} and $\phi\sim \phi+2\pi$.
The connected part of the group ${SO}_c(2,2)$ is $SL(2,\mathbb{R})\otimes SL(2,\mathbb{R})/\mathbb{Z}_2$. The group elements $(g_L,g_R)\in {SO}_c(2,2)$ act on a point $p$ according to
\be
p\rightarrow g_Lpg_R^t.
\ee
From this, we see that the $\mathbb{Z}_2$ symmetry correspond to the equivalence relation $(g_L,g_R)\sim (-g_L,-g_R)$.
A convenient basis of generators $\{J_1,J_2,J_3\}\times \{\tilde{J}_1,\tilde{J}_2,\tilde{J}_3\}$ of the isometry group $SL(2,\mathbb{R})\times SL(2,\mathbb{R})$ is
\begin{equation}
\begin{aligned}
J_{1} & \equiv-\frac{1}{2}\left(J_{X U}-J_{Y V}\right), \quad \tilde{J}_{1} \equiv-\frac{1}{2}\left(J_{X U}+J_{Y V}\right)\\
J_{2} &\equiv -\frac{1}{2}\left(J_{YU}+J_{XV}\right),\quad  \tilde{J}_{2} \equiv-\frac{1}{2}\left(J_{Y U}-J_{X V}\right) \\
J_{3} &\equiv-\frac{1}{2}\left(J_{UV}-J_{XY}\right), \quad \tilde{J}_{3} \equiv \frac{1}{2}\left(J_{UV}+J_{XY}\right)
\end{aligned}
\end{equation}
where the Killing vectors $J_{ab}=\bar{x}_a\bar{\partial} _{b}-\bar{x}_b\bar{\partial} _{a}$ obey the $SO(2,2)$ algebra
\be
[J_{ab},J_{cd}]=\bar{\eta}_{ac}J_{bd}-\bar{\eta}_{ad}J_{bc}-\bar{\eta}_{bc}J_{ad}+\bar{\eta}_{bd}J_{ac}
\ee
In matrix representation, the generators are expressed as
\begin{equation}
J_{1}=-\frac{1}{2} \gamma_{1}, \quad J_{2}=-\frac{1}{2} \gamma_{2}, \quad J_{3}=-\frac{1}{2} \gamma_{3}
\end{equation}
where\footnote{Our matrix representation of $p$ is different from  that defined in \cite{Aminneborg:1998si,Krasnov:2001va}, which causes the generators to be slightly different.}
\begin{equation}
\gamma_{1}=\left(\begin{array}{cc}
1 & 0 \\
0 & -1
\end{array}\right), \quad \gamma_{2}=\left(\begin{array}{cc}
0 & 1 \\
1 & 0
\end{array}\right),\quad
\gamma_{3}=\left(\begin{array}{cc}
0 & 1 \\
-1 & 0
\end{array}\right)
\end{equation}
and similarly for $\tilde{J}_i$\footnote{In matrix representation, $\tilde{J}_i$ takes the same matrix form as $J_i=-\frac{1}{2}\gamma_i$ but the infinitesimal transformations on $p$ are different from those of $J_i$'s, since $J_i: p\rightarrow -\frac{1}{2}\gamma_i p$ while $\tilde{J}_i:p\rightarrow -\frac{1}{2}p\gamma_i^t$.}.

\begin{figure}
\centering
\begin{tikzpicture}
\draw[black,very thick,solid] (0,0) -- (0,7.2);
\draw[black,very thick,solid] (-3.6,0) -- (-3.6,7.2);

\draw[black,very thin,dashed] (-3.6,1.8) -- (-1.8,3.6);
\draw[black,very thin,dashed] (0,1.8) -- (-1.8,3.6);
\draw[black,very thin,dashed] (-3.6,1.8) -- (-1.8,0);
\draw[black,very thin,dashed] (0,1.8) -- (-1.8,0);

\draw[black,very thin,dashed] (-3.6,5.4) -- (-1.8,7.2);
\draw[black,very thin,dashed] (0,5.4) -- (-1.8,7.2);
\draw[black,very thin,dashed] (-3.6,5.4) -- (-1.8,3.6);
\draw[black,very thin,dashed] (0,5.4) -- (-1.8,3.6);

\draw[black,thin,dotted] (-1.8,0) -- (-1.8,7.2);
\draw[black,thin,dotted] (-3.6,3.6) -- (0,3.6);

\draw[black,thin,dotted] plot[variable=\t,samples=10,domain=0:63] ({0.89*tan(\t)-3.6},{0.81*sec(\t)+1.8});
\draw[black,thin,dotted] plot[variable=\t,samples=10,domain=-63:0] ({0.89*tan(\t)},{0.81*sec(\t)+1.8});

\draw[black,thin,dotted] plot[variable=\t,samples=10,domain=0:63] ({0.89*tan(\t)-3.6},{-0.81*sec(\t)+5.4});
\draw[black,thin,dotted] plot[variable=\t,samples=10,domain=-63:0] ({0.89*tan(\t)},{-0.81*sec(\t)+5.4});

\draw[black,thin,dotted] plot[variable=\t,samples=10,domain=0:63] ({0.89*tan(\t)-3.6},{-0.81*sec(\t)+1.8});
\draw[black,thin,dotted] plot[variable=\t,samples=10,domain=-63:0] ({0.89*tan(\t)},{-0.81*sec(\t)+1.8});

\draw[black,thin,dotted] plot[variable=\t,samples=10,domain=0:63] ({0.89*tan(\t)-3.6},{0.81*sec(\t)+5.4});
\draw[black,thin,dotted] plot[variable=\t,samples=10,domain=-63:0] ({0.89*tan(\t)},{0.81*sec(\t)+5.4});

\draw[black,thin,dotted] plot[variable=\t,samples=10,domain=-63:63] ({-0.81*sec(\t)},{0.89*tan(\t)+5.4});
\draw[black,thin,dotted] plot[variable=\t,samples=10,domain=-63:63] ({0.81*sec(\t)-3.6},{0.89*tan(\t)+5.4});

\draw[black,thin,dotted] plot[variable=\t,samples=10,domain=-63:63] ({-0.81*sec(\t)},{0.89*tan(\t)+1.8});
\draw[black,thin,dotted] plot[variable=\t,samples=10,domain=-63:63] ({0.81*sec(\t)-3.6},{0.89*tan(\t)+1.8});

\node[text width=1.5cm] at (0.3,-0.3) {$\phi=\pi$};
\node[text width=1.5cm] at (-3.3,-0.3) {$\phi=-\pi$};

\node[text width=1.5cm] at (-4.1,7.1) {$t=\pi$};
\node[text width=1.5cm] at (-4.1,0.3) {$t=-\pi$};

\node[text width=1.5cm] at (-0.3,3.8) {$I$};
\node[text width=1.5cm] at (-0.3,7.1) {$II$};
\node[text width=1.5cm] at (-0.8,5.4) {$III$};
\node[text width=1.5cm] at (-0.8,1.8) {$IV$};

\filldraw[black] (-1.8,3.6) circle (1pt) node[anchor=east] {$e$};
\end{tikzpicture}
\caption{The group manifold of $SL(2,\mathbb{R})$, which is also the Penrose diagram of AdS$_3$. The dotted lines represent the action of the group elements of $SL(2,\mathbb{R})$ on the identity element $e$ placed at the origin of AdS$_3$ in global coordinates. The isometries of $SL(2,\mathbb{R})$ are classified depending on which region the element $e$ is mapped to. Dashed lines represent null rays.}\label{fig:AdS3}
\end{figure}
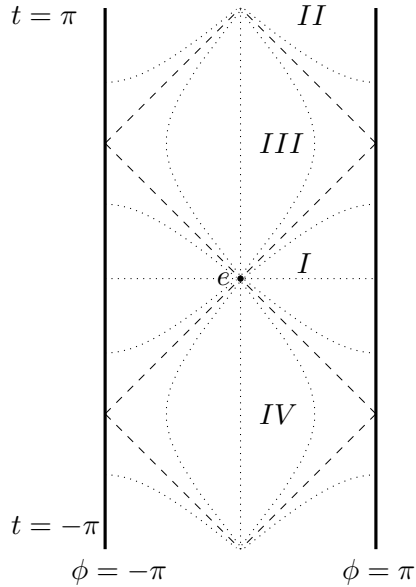

To understand the action of the group elements $(g_L,g_R)$, we will describe AdS$_3$ as the group manifold of $SL(2,\mathbb{R})$,  with the Penrose diagram shown in figure \ref{fig:AdS3}. The action of group elements $g\in SL(2,\mathbb{R})$ on the identity element $e$ is shown there, according to which they are classified into conjugacy classes depending on where the point $e\rightarrow geg^{t}=gg^t$ lies,
\begin{center}
\begin{tabular}{ c c c c c c}
 Hyperbolic && $\Tr g>2$ && $gg^t\in I$ \\
 Hyperbolic && $\Tr g<-2$ && $gg^t\in II$ \\
 Elliptic && $|\Tr g|<2$ && $gg^t\in III,IV$ \\
 Parabolic && $|\Tr g|=2$ && $gg^t\in$ light cones
\end{tabular}
\end{center}

We will focus on the action of subgroups $\Gamma\subseteq {SO}_c(2,2)$ with $\Tr g>2$ hyperbolic elements, whose fixed points are on the boundary of AdS$_3$. This is because it ensures that $AdS_3/\Gamma$ is free of conical singularities and closed timelike curves \cite{Aminneborg:1997pz, Brill:1998pr}. Removing from the spacetime the past and future of those fixed points
yields the restricted spacetime $\widehat{\text{AdS}_3}$ where the action of the quotient on the spacetime is free of pathologies and leads to a spacetime $\widehat{\text{AdS}_3}/\Gamma$. We will illustrate this process by reviewing the construction of $\widehat{\text{AdS}_3}/\Gamma$ in the case of BTZ black holes \cite{Banados:1992wn,Banados:1992gq} and three-boundary black holes \cite{Brill:1995jv,Aminneborg:1997pz, Brill:1998pr}. We also discuss generalizations to $n$-boundary black holes with and without non-trivial topologies \cite{Aminneborg:1997pz, Brill:1998pr,Aminneborg:1998si}. A Cauchy slice of these geometries is a Riemannian manifold of genus $g$ and boundary number $n$. So, we can classify the black hole geometries by a 2-tuple $(n,g)$. In the non-rotating case, the number of parameters (or in other words, dimension of the moduli space) 
needed to specify the $(n,g)$ geometry is equal to $1$ for $(2,0)$ and is $6g-6+3n$ otherwise. In the rotating case, this number is doubled.

Before reviewing the construction of these geometries, we will give general formulas for calculating the geodesic distance. The group manifold representation allow us to easily calculate the geodesic distances $d(p,q)$ between two arbitrary points, $p$ and $q$ \cite{Maxfield:2014kra}. In particular, if $p$ and $q$ are connected by a spacelike geodesic, then
\be\label{Lspacelike}
d(p,q)=\cosh^{-1}\left(\frac{\Tr\left(p^{-1}q\right)}{2}\right).
\ee
With a timelike geodesic connecting $p$ and $q$, the geodesic distance is
\be\label{Ltimelike}
d(p,q)=\cos^{-1}\left(\frac{\Tr\left(p^{-1}q\right)}{2}\right).
\ee
When $\Tr\left(p^{-1}q\right)<-2$, there is no geodesic connecting $p$ and $q$.


We now discuss various cases in detail.

\subsection*{BTZ black hole}
\begin{figure}[t]
\centering
\begin{tikzpicture}
\draw[black,thick,solid] (0,0) circle[radius=3];
\draw [blue] (-2.5,-1.6) to[out=30,in=150] (2.5,-1.6);
\draw [blue] (-2.5,1.6) to[out=-30,in=-150] (2.5,1.6);

\filldraw[black] (0,3) circle (1.5pt) node[anchor=south] {};
\filldraw[black] (0,-3) circle (1.5pt) node[anchor=south] {};
\draw[black,thin,dashed] (0,-.88) -- (0,0.88);
\node[text width=0.8cm] at (0.55,0) {$H$};
\node[blue,text width=0.1cm] at (0.55,.95) {/};
\node[blue,text width=0.1cm] at (0.55,-.95) {/};
\end{tikzpicture}
\caption{A Cauchy slice of a BTZ black hole shown as a quotient of AdS$_3$. The action of $\gamma$ identifies the two blue geodesics, and the region between them is the fundamental domain of the quotient. The minimal geodesic $H$ separating the two coincides with the event horizon of the black hole. In the non-rotating case, this slice is at $t=0$. But in the case of rotation, there is a relative boost between the two identified geodesics.}\label{fig:BTZ}
\end{figure}
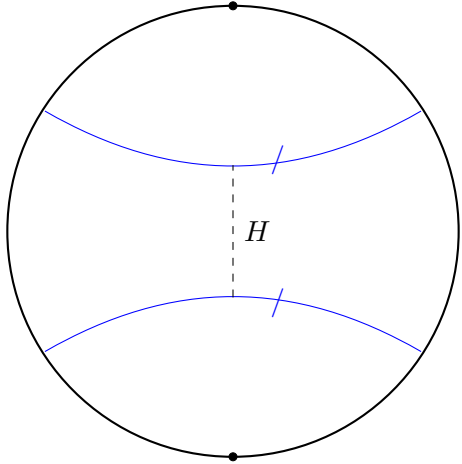

In this case, the subgroup $\Gamma$ is generated by a single element
\be\label{BTZgamma}
\gamma_{BTZ}=(g_{L,BTZ},g_{R,BTZ})=\left(e^{\ell \xi_{L,BTZ}},e^{\tilde{\ell} \xi_{R,BTZ}}\right)
\ee
and a convenient choice for $\xi_{L,BTZ}$ and $\xi_{R,BTZ}$ is
\begin{equation}
\xi_{L,BTZ}=-J_2,\quad \xi_{R,BTZ}=-\tilde{J}_2
\end{equation}
with $\ell=2\pi(r_+ +r_-)$ and $\tilde{\ell}=2\pi(r_+ -r_-)$ being two positive real parameters. 
In matrix representation, this gives
\be
g_{L,BTZ}=\begin{pmatrix}
\cosh \left(\frac{\ell}{2}\right)& \sinh \left(\frac{\ell}{2}\right) \\
\sinh \left(\frac{\ell}{2}\right)& \cosh \left(\frac{\ell}{2}\right)
\end{pmatrix},\quad
g_{R,BTZ}=\begin{pmatrix}
\cosh \left(\frac{\tilde{\ell}}{2}\right)& \sinh \left(\frac{\tilde{\ell}}{2}\right) \\
\sinh \left(\frac{\tilde{\ell}}{2}\right)& \cosh \left(\frac{\tilde{\ell}}{2}\right)
\end{pmatrix}.
\ee
The isometry $\gamma$ has two fixed points at the boundary given by $t=0,\phi=\pi/2$ and $t=0,\phi=3\pi/2$. Removing the past and future regions of these fixed points gives the restricted space $\widehat{\text{AdS}_3}$. Any two geodesics that are related by the isometry $\gamma_{BTZ}$ are identified, and we can choose a region that is bounded by such a pair of geodesics as the fundamental domain of $\widehat{\text{AdS}_3}/\Gamma$, see figure \ref{fig:BTZ}. The minimal length between these two geodesics is uniquely determined by $r_+$ and $r_-$, and is the intersection of the geodesic connecting the fixed points with the fundamental domain. This defines the two-sided BTZ black hole, where each side is covered by the usual BTZ coordinates
\be\label{BTZmetric}
ds^2=-\frac{\left(r_B^{2}-r_{+}^{2}\right)\left(r_B^{2}-r_{-}^{2}\right)}{r_B^{2}} d t_B^{2}+\frac{r_B^{2}}{\left(r_B^{2}-r_{+}^{2}\right)\left(r_B^{2}-r_{-}^{2}\right)} d r_B^{2}+r_B^2{\left(d\phi_B-\frac{r_+r_-}{r_B^2}dt_B\right)}^2
\ee
where the subscript B means that we are using BTZ coordinates.
The thermodynamic quantities related to the black hole are
\begin{equation}
\label{BTZthermo}
\begin{aligned}
M&=\frac{r_{+}^{2}+r_{-}^{2}}{8G_N}=\frac{\ell^2+\tilde{\ell}^2}{64\pi^2 G_N}, \quad J=\frac{ r_{+} r_{-}}{4G_N}=\frac{\ell^2-\tilde{\ell}^2}{64\pi^2 G_N}\\
T_{H}&=\frac{1}{\beta}=\frac{r_{+}^{2}-r_{-}^{2}}{2 \pi r_{+}}=\frac{\ell \tilde{\ell}}{2\pi^2(\ell+\tilde{\ell})},  \quad \Omega_H=\frac{r_{-}}{r_{+}}=\frac{\ell-\tilde{\ell}}{\ell+\tilde{\ell}}.
\end{aligned}
\end{equation}

By writing the point $p$ in \eqref{pembedd} in terms of the BTZ coordinates using the transformation
\begin{equation}
\label{BTZembed}
\begin{aligned}
U&= \sqrt{\frac{r_B^2-r_-^2}{r_+^2-r_-^2}} \cosh \left({r_{+}} \phi_B+{r_{-}} t_B\right), \quad X= \sqrt{\frac{r_B^2-r_+^2}{r_+^2-r_-^2}} \cosh \left({r_{+}} t_B+{r_{-}} \phi_B\right),   \\
V&= \sqrt{\frac{r_B^2-r_+^2}{r_+^2-r_-^2}} \sinh \left({r_{+}} t_B+{r_{-}} \phi_B \right), \quad Y=\sqrt{\frac{r_B^2-r_-^2}{r_+^2-r_-^2}} \sinh \left({r_{+}} \phi_B+{r_{-}} t_B\right)
\end{aligned}
\end{equation}
one can show that the action of $\gamma_{BTZ}$ on $p$ is simply to map $\phi_B \rightarrow\phi_B+2\pi$. The length of the bifurcation surface (horizon length) generated by $\gamma$ can be found from \eqref{Lspacelike} to be \cite{Maxfield:2014kra}
\be\label{horizonLength}
h=\cosh^{-1}\left(\frac{\Tr g_{L,BTZ}}{2}\right)+\cosh^{-1}\left(\frac{\Tr g_{R,BTZ}}{2}\right)
\ee
From \eqref{BTZgamma}, we see that this gives the expected horizon length of $\frac{\ell+\tilde{\ell}}{2}=2\pi r_+$.
\subsection*{Three-boundary black hole}
The subgroup $\Gamma$ in this case is generated by two elements $\gamma_i=(g_{iL},g_{iR}), i=1,2$. We choose the first one to be the same as the isometry used to construct the BTZ black hole\footnote{Note that, here, the choice of generators $\gamma_i$ is not unique. Other choices could be used, as long as they fall in certain conjugacy classes. Our choice here is convenient for calculation, but as we will see, it defines a conformal frame in which the third boundary region becomes vanishingly small in the hot limit. In appendix \ref{appSymm}, we give an example of another construction of the same geometry and discuss how it differs from the one used here.}
\be
\gamma_1=(g_{1L},g_{1R})=(e^{\ell_1\xi_{1L}},e^{\tilde{\ell}_1\xi_{1R}})
\ee
where $\xi_{1L}=-J_2$ and $\xi_{1R}=-\tilde{J}_2$.
\begin{figure}[t]
\centering
\begin{tikzpicture}
\draw[black,thick,solid] (0,0) circle (3cm);
\draw[black,thin,dashed] (0,-.99) -- (0,0.99);
\node[text width=0.8cm] at (0.55,0) {$H_1$};
\draw [dashed] (-2.88,0.85) to[out=-16.6,in=16.6] (-2.88,-0.85);
\draw [dashed] (-2.78,1.12) to [out=-21.8,in=-40](-2.28,1.94);
\draw [dashed] (-2.78,-1.12) to [out=21.8,in=40](-2.28,-1.94);
\draw [blue,thick] (-2.36,-1.85) to[out=38,in=142] (2.36,-1.85);
\path [fill=white] (-2.36,-1.85) to[out=38,in=142] (2.36,-1.85) to (0,-3) to (-2.36,-1.85);
\draw [blue,thick] (-2.36,1.85) to[out=-38,in=-142] (2.36,1.85);
\path [fill=white] (-2.36,1.85) to[out=-38,in=-142] (2.36,1.85) to (0,3) to (-2.36,1.85);
\draw [thick,red] (-2.95,0.52) to[out=-10,in=-25] (-2.72,1.27);
\path [fill=white] (-2.95,0.52) to[out=-10,in=-25] (-2.72,1.27) to (-2.95,0.52);
\draw [thick,red] (-2.95,-0.52) to[out=10,in=25] (-2.72,-1.27);
\path [fill=white] (-2.95,-0.52) to[out=10,in=25] (-2.72,-1.27) to (-2.95,-0.52);
\node[blue,text width=0.1cm] at (0.55,1.05) {/};
\node[blue,text width=0.1cm] at (0.55,-1.05) {/};
\node[red] at (-2.8, 0.602944) {//};
\node[red] at (-2.8, -0.602944) {//};
\node[text width=0.8cm] at (-1.9,0) {$H_2$};
\filldraw[black] (0,3) circle (1.5pt) node[anchor=south] {};
\filldraw[black] (0,-3) circle (1.5pt) node[anchor=south] {};
\node[text width=0.8cm] at (-1.71433, 1.03008) {$H_3'$};
\node[text width=0.8cm] at (-1.71433, -1.03008) {$H_3''$};
\filldraw[black] (-2.87646, 0.852046) circle (1.5pt) node[anchor=south] {};
\filldraw[black] (-2.87646, -0.852046) circle (1.5pt) node[anchor=south] {};
\filldraw[black] (-2.78155, 1.12382) circle (1.5pt) node[anchor=south] {};
\filldraw[black] (-2.2,2.05) circle (1.5pt) node[anchor=south] {};
\filldraw[black] (-2.78155, -1.12382) circle (1.5pt) node[anchor=south] {};
\filldraw[black] (-2.2, -2.05) circle (1.5pt) node[anchor=south] {};
\end{tikzpicture}
\caption{A Cauchy slice of the three-boundary black hole shown as a quotient of AdS$_3$. The action of $\gamma_1$ identifies the two blue geodesics while $\gamma_2$ identifies the two red geodesics. The event horizons of the three boundaries $H_1$, $H_2$, and $H_3=H_3'\cup H_3''$ are also shown, where each of them coincide with the geodesic connecting the fixed points of the isometries $\gamma_1$, $\gamma_2$, and $\gamma_3$, respectively. Note that $\gamma_3$ has four fixed points instead of two, because it defines the third asymptotic region as the union of two separate regions in the Cauchy slice. In the case of no rotation, this slice is that of $t=0$.}\label{fig:3bdy}
\end{figure}
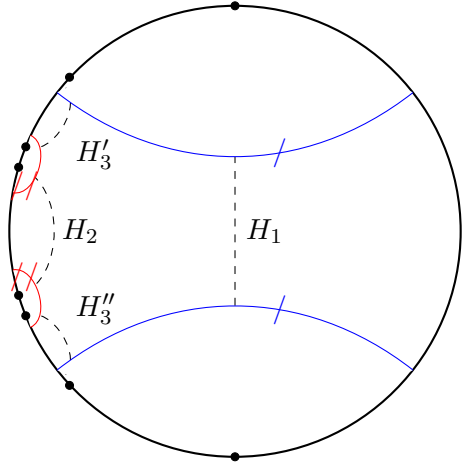
The second element is given by
\be
\gamma_2=(g_{2L},g_{2R})=(e^{\ell_2\xi_{2L}},e^{\tilde{\ell}_2\xi_{2R}})
\ee
where $\xi_{2L}=-(J_2\cosh\alpha+J_3\sinh\alpha)$ and $\xi_{2R}=-(\tilde{J}_2\cosh\tilde{\alpha}+\tilde{J}_3\sinh\tilde{\alpha})$. In matrix representation, this is
\be
g_{2L}=\begin{pmatrix}
\cosh \left(\frac{\ell_{2}}{2}\right)& e^\alpha\sinh \left(\frac{\ell_{2}}{2}\right) \\
e^{-\alpha}\sinh \left(\frac{\ell_{2}}{2}\right)& \cosh \left(\frac{\ell_{2}}{2}\right)
\end{pmatrix},\quad
g_{2R}=\begin{pmatrix}
\cosh \left(\frac{\tilde{\ell}_{2}}{2}\right)& e^{\tilde{\alpha}}\sinh \left(\frac{\tilde{\ell}_{2}}{2}\right) \\
e^{-\tilde{\alpha}}\sinh \left(\frac{\tilde{\ell}_{2}}{2}\right)& \cosh \left(\frac{\tilde{\ell}_{2}}{2}\right)
\end{pmatrix}.
\ee
These two isometries define the first and second asymptotic regions, with the event horizons of these regions lying along the geodesics connecting the fixed points of $\gamma_1$ and $\gamma_2$, respectively.

The isometries that define the third asymptotic region are not independent of the above two. They are $\gamma_3'=-\gamma_1\gamma_2^{-1}\Rightarrow (g_{3L}',g_{3R}')=(-g_{1L}g_{2L}^{-1},-g_{1R}g_{2R}^{-1})$ and $\gamma_3''=-\gamma_1^{-1}\gamma_2\Rightarrow (g_{3L}'',g_{3R}'')=(-g_{1L}^{-1}g_{2L},-g_{1R}^{-1}g_{2R})$\footnote{Although $\gamma_3'$ and $\gamma_3''$ are both isometries defining the third region, for simplicity of notation, later we will refer to them collectively as $\gamma_3$.}, corresponding to the two parts of the third boundary region as seen from the covering space.
The resulting spacetime is a black hole with three asymptotic boundaries, as shown in figure \ref{fig:3bdy}. The spacetime in each asymptotic region is isometric to the exterior region of a BTZ black hole. Hence, each asymptotic region can be covered by the same metric \eqref{BTZmetric} for $r_B>r_+$. The lengths of the horizons generated by these isometries can be found from \eqref{horizonLength} to be
\be\label{horizonlengths}
h_1=\frac{\ell_1+\tilde{\ell}_1}{2},\quad h_2=\frac{\ell_2+\tilde{\ell}_2}{2},\quad\text{and}\quad h_3=\frac{\ell_3+\tilde{\ell}_3}{2},
\ee
where we have defined
\be
\ell_3\equiv 2\cosh^{-1}\left(\frac{\Tr g_{3L}}{2}\right),\quad\text{and}\quad \tilde{\ell}_3\equiv 2\cosh^{-1}\left(\frac{\Tr g_{3R}}{2}\right).
\ee
The parameter $\alpha$ can in turn be expressed using $\ell_i, i=1,2,3$:
\begin{equation}\label{alphaEq}
\begin{aligned}
\cosh \alpha=\frac{\cosh \frac{\ell_3}{2}+\cosh \frac{\ell_1}{2}\cosh\frac{\ell_2}{2}}{\sinh\frac{\ell_1}{2}\sinh\frac{\ell_2}{2}},
\end{aligned}
\end{equation}
and similarly for $\tilde{\alpha}$. Each asymptotic region can be associated with independent thermodynamic parameters \eqref{BTZthermo}. The angular velocity associated to a horizon generated by an isometry $\gamma_i$ can be given in terms of the isometry elements as \cite{Krasnov:2001va}
\be
\Omega_i=\frac{\cosh^{-1}\left(\frac{\Tr g_{iL}}{2}\right)-\cosh^{-1}\left(\frac{\Tr g_{iR}}{2}\right)}{\cosh^{-1}\left(\frac{\Tr g_{iL}}{2}\right)+\cosh^{-1}\left(\frac{\Tr g_{iR}}{2}\right)},
\ee
which gives
\be
\Omega_1=\frac{\ell_1-\tilde{\ell}_1}{\ell_1+\tilde{\ell}_1},\quad \Omega_2=\frac{\ell_2-\tilde{\ell}_2}{\ell_2+\tilde{\ell}_2},\quad\text{and}\quad\Omega_3=\frac{\ell_3-\tilde{\ell}_3}{\ell_3+\tilde{\ell}_3}
\ee
for the three boundaries. From this and the fact that the horizon lengths $h_i$ are given by $2\pi r_{+,i}$, we can relate the geometric parameters $\ell_i$ and $\tilde{\ell}_i$ for each boundary to the inner and outer horizon lengths of the corresponding black hole.  The resulting relation is
\be
r_{\pm,i}=\frac{\ell_i\pm\tilde{\ell}_i}{4\pi}
\ee
for $i=1,2,3$. We see that setting $\tilde{\ell}_i=0$ corresponds to the extremal case\footnote{Here we have implicitly chosen a direction of spinning. For the other choice, $\ell_i=0$ would correspond to an extremal black hole.}, while setting $\ell_i=\tilde{\ell}_i$ corresponds to the non-rotating case. The unique feature of $(3,0)$ geometry (and any geometry $(n,g)$ other than BTZ) is the existence of a region between the horizons $H_1$, $H_2$, and $H_3$ that does not intersect the causal past and future of any asymptotic region. This region is called the causal shadow of the spacetime \cite{Headrick:2014cta}, and it will be important in our discussion of traversability below. The causal shadow region is bounded by closed geodesics, which allow us to calculate its area using the Gauss-Bonnet theorem, giving $A_\text{CS}=2(n-2+2g)\pi$ for general $(n,g)$ spacetimes \cite{Marolf:2015vma}. This shows that the causal shadow region exists for all geometries except $(2,0)$.

\subsection*{General $(n,g)$ black holes}
More general black hole geometries can be constructed following the same method as discussed above.  For the case without rotations, general $(n,g)$ geometries could be constructed using a cut-and-paste procedure \cite{Aminneborg:1997pz, Brill:1998pr}, and this could be easily generalized to cases with rotations, as we review below.

The simplest way to see this is to note that any $(n,g)$ black hole can be constructed from $2g+n-2$ copies of the $(3,0)$ geometry (so-called ``pair-of-pants" geometry) through a process of cutting, twisting, and gluing. Since the $(3,0)$ geometry is everywhere locally AdS$_3$, the geometry that results from a process of cutting, twisting, and gluing different copies of it is also locally AdS$_3$ and, therefore, is a solution of Einstein gravity. We will illustrate this process in the case of $n$ asymptotic regions and in case of genus $g$.

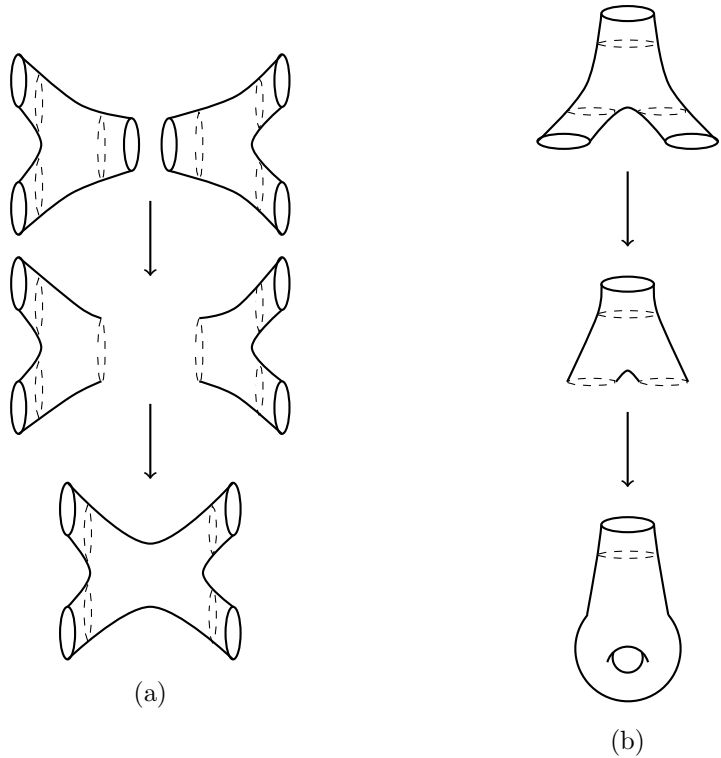
\begin{figure}[t]\centering
\begin{subfigure}{0.4\textwidth}\centering
\begin{tikzpicture}
\draw[black,thick,solid] (-2,2.1) ellipse (0.1 and 0.35);
\draw[black,thick,solid] (-2,0.4) ellipse (0.1 and 0.35);
\draw[black,thick,solid] (-0.5,1.25) ellipse (0.1 and 0.35);
\draw [black,thick,solid] plot [smooth, tension=0.6] coordinates { (-0.5,0.9) (-1.3,0.6) (-2.0,0.05)};
\draw [black,thick,solid] plot [smooth, tension=0.6] coordinates { (-2.0,0.75) (-1.7,1.25) (-2.0,1.75)};
\draw [black,thick,solid] plot [smooth, tension=0.6] coordinates { (-2.0,2.45) (-1.2,1.79) (-0.5,1.6)};
\draw[black,thin,dashed] plot[variable=\t,samples=10,domain=-170:170] ({(0.05*cos(\t))*cos(-0)-(0.37*sin(\t))*sin(-0)-1.73},{(0.05*cos(\t))*sin(-0)+(0.37*sin(\t))*cos(-0)+1.8});
\draw[black,thin,dashed] plot[variable=\t,samples=10,domain=-170:170] ({(0.05*cos(\t))*cos(-0)-(0.37*sin(\t))*sin(-0)-1.73},{(0.05*cos(\t))*sin(-0)+(0.37*sin(\t))*cos(-0)+0.69});
\draw[black,thin,dashed] plot[variable=\t,samples=10,domain=-170:170] ({(0.05*cos(\t))*cos(-0)-(0.44*sin(\t))*sin(-0)-0.9},{(0.05*cos(\t))*sin(-0)+(0.44*sin(\t))*cos(-0)+1.2});

\draw[black,thick,solid] (1.5,2.1) ellipse (0.1 and 0.35);
\draw[black,thick,solid] (1.5,0.4) ellipse (0.1 and 0.35);
\draw[black,thick,solid] (0.0,1.25) ellipse (0.1 and 0.35);
\draw [black,thick,solid] plot [smooth, tension=0.6] coordinates { (0.0,0.9) (0.9,0.6) (1.5,0.05)};
\draw [black,thick,solid] plot [smooth, tension=0.6] coordinates { (1.5,0.75) (1.1,1.25) (1.5,1.75)};
\draw [black,thick,solid] plot [smooth, tension=0.6] coordinates { (1.5,2.45) (0.9,1.9) (0.0,1.6)};
\draw[black,thin,dashed] plot[variable=\t,samples=10,domain=-170:170] ({(0.05*cos(\t))*cos(-0)-(0.33*sin(\t))*sin(-0)+1.2},{(0.05*cos(\t))*sin(-0)+(0.33*sin(\t))*cos(-0)+1.8});
\draw[black,thin,dashed] plot[variable=\t,samples=10,domain=-170:170] ({(0.05*cos(\t))*cos(-0)-(0.33*sin(\t))*sin(-0)+1.2},{(0.05*cos(\t))*sin(-0)+(0.33*sin(\t))*cos(-0)+0.69});
\draw[black,thin,dashed] plot[variable=\t,samples=10,domain=-170:170] ({(0.05*cos(\t))*cos(-0)-(0.45*sin(\t))*sin(-0)+0.4},{(0.05*cos(\t))*sin(-0)+(0.45*sin(\t))*cos(-0)+1.23});

\draw[black,thick,solid,->] (-0.25,0.5) -- (-0.25,-0.5);

\draw[black,thick,solid] (-2,-0.6) ellipse (0.1 and 0.35);
\draw[black,thick,solid] (-2,-2.25) ellipse (0.1 and 0.35);
\draw [black,thick,solid] plot [smooth, tension=0.6] coordinates { (-0.9,-1.9) (-1.3,-2.1) (-2.0,-2.6)};
\draw [black,thick,solid] plot [smooth, tension=0.6] coordinates { (-2.0,-1.9) (-1.7,-1.45) (-2.0,-0.95)};
\draw [black,thick,solid] plot [smooth, tension=0.6] coordinates { (-2.0,-0.25) (-1.2,-0.91) (-0.9,-1.06)};
\draw[black,thin,dashed] plot[variable=\t,samples=10,domain=-170:170] ({(0.05*cos(\t))*cos(-0)-(0.37*sin(\t))*sin(-0)-1.73},{(0.05*cos(\t))*sin(-0)+(0.37*sin(\t))*cos(-0)-0.9});
\draw[black,thin,dashed] plot[variable=\t,samples=10,domain=-170:170] ({(0.05*cos(\t))*cos(-0)-(0.37*sin(\t))*sin(-0)-1.73},{(0.05*cos(\t))*sin(-0)+(0.37*sin(\t))*cos(-0)-2.01});
\draw[black,thin,dashed] plot[variable=\t,samples=10,domain=-170:170] ({(0.05*cos(\t))*cos(-0)-(0.44*sin(\t))*sin(-0)-0.9},{(0.05*cos(\t))*sin(-0)+(0.44*sin(\t))*cos(-0)-1.5});

\draw[black,thick,solid] (1.5,-0.6) ellipse (0.1 and 0.35);
\draw[black,thick,solid] (1.5,-2.25) ellipse (0.1 and 0.35);
\draw [black,thick,solid] plot [smooth, tension=0.6] coordinates { (0.4,-1.9) (0.9,-2.1) (1.5,-2.6)};
\draw [black,thick,solid] plot [smooth, tension=0.6] coordinates { (1.5,-1.9) (1.1,-1.45) (1.5,-0.95)};
\draw [black,thick,solid] plot [smooth, tension=0.6] coordinates { (1.5,-0.25) (0.9,-0.91) (0.4,-1.06)};
\draw[black,thin,dashed] plot[variable=\t,samples=10,domain=-170:170] ({(0.05*cos(\t))*cos(-0)-(0.37*sin(\t))*sin(-0)+1.2},{(0.05*cos(\t))*sin(-0)+(0.34*sin(\t))*cos(-0)-0.9});
\draw[black,thin,dashed] plot[variable=\t,samples=10,domain=-170:170] ({(0.05*cos(\t))*cos(-0)-(0.37*sin(\t))*sin(-0)+1.2},{(0.05*cos(\t))*sin(-0)+(0.34*sin(\t))*cos(-0)-2.01});
\draw[black,thin,dashed] plot[variable=\t,samples=10,domain=-170:170] ({(0.05*cos(\t))*cos(-0)-(0.44*sin(\t))*sin(-0)+0.4},{(0.05*cos(\t))*sin(-0)+(0.44*sin(\t))*cos(-0)-1.5});

\draw[black,thick,solid,->] (-0.25,-2.2) -- (-0.25,-3.2);

\draw[black,thick,solid] (-1.35,-3.6) ellipse (0.1 and 0.35);
\draw[black,thick,solid] (-1.35,-5.25) ellipse (0.1 and 0.35);
\draw [black,thick,solid] plot [smooth, tension=0.6] coordinates { (-1.35,-4.9) (-1.05,-4.45) (-1.35,-3.95)};
\draw[black,thin,dashed] plot[variable=\t,samples=10,domain=-170:170] ({(0.05*cos(\t))*cos(-0)-(0.37*sin(\t))*sin(-0)-1.08},{(0.05*cos(\t))*sin(-0)+(0.37*sin(\t))*cos(-0)-3.92});
\draw[black,thin,dashed] plot[variable=\t,samples=10,domain=-170:170] ({(0.05*cos(\t))*cos(-0)-(0.37*sin(\t))*sin(-0)-1.08},{(0.05*cos(\t))*sin(-0)+(0.37*sin(\t))*cos(-0)-4.98});

\draw[black,thick,solid] (0.85,-3.6) ellipse (0.1 and 0.35);
\draw[black,thick,solid] (0.85,-5.25) ellipse (0.1 and 0.35);
\draw [black,thick,solid] plot [smooth, tension=0.6] coordinates { (-1.35,-5.6)  (-0.25,-4.9) (0.85,-5.6)};
\draw [black,thick,solid] plot [smooth, tension=0.6] coordinates { (0.85,-4.9) (0.45,-4.45) (0.85,-3.95)};
\draw [black,thick,solid] plot [smooth, tension=0.6] coordinates { (-1.35,-3.25) (-0.25,-4.06) (0.85,-3.25)};
\draw[black,thin,dashed] plot[variable=\t,samples=10,domain=-170:170] ({(0.05*cos(\t))*cos(-0)-(0.37*sin(\t))*sin(-0)+0.58},{(0.05*cos(\t))*sin(-0)+(0.32*sin(\t))*cos(-0)-3.88});
\draw[black,thin,dashed] plot[variable=\t,samples=10,domain=-170:170] ({(0.05*cos(\t))*cos(-0)-(0.37*sin(\t))*sin(-0)+0.58},{(0.05*cos(\t))*sin(-0)+(0.36*sin(\t))*cos(-0)-5.02});
\end{tikzpicture}
 \caption{}
\label{fig:4bdy}
\end{subfigure}
\begin{subfigure}{0.4\textwidth}\centering
\begin{tikzpicture}
\draw[black,thick,solid] (0,2.1) ellipse (0.35 and 0.1);
\draw[black,thick,solid] (0.85,0.4) ellipse (0.35 and 0.1);
\draw[black,thick,solid] (-0.85,0.4) ellipse (0.35 and 0.1);
\draw [black,thick,solid] plot [smooth, tension=0.6] coordinates { (-1.2,0.4)  (-0.55,1.15) (-0.35,2.1)};
\draw [black,thick,solid] plot [smooth, tension=0.6] coordinates { (0.35,2.1)  (0.55,1.15) (1.2,0.4)};
\draw [black,thick,solid] plot [smooth, tension=0.6] coordinates { (-0.5,0.4)  (0.0,0.85) (0.5,0.4)};
\draw[black,thin,dashed] plot[variable=\t,samples=10,domain=-170:170] ({(0.4*cos(\t))*cos(-0)-(0.05*sin(\t))*sin(-0)+0.0},{(0.4*cos(\t))*sin(-0)+(0.05*sin(\t))*cos(-0)+1.7});
\draw[black,thin,dashed] plot[variable=\t,samples=10,domain=-170:170] ({(0.325*cos(\t))*cos(-0)-(0.05*sin(\t))*sin(-0)-0.48},{(0.325*cos(\t))*sin(-0)+(0.05*sin(\t))*cos(-0)+0.8});
\draw[black,thin,dashed] plot[variable=\t,samples=10,domain=-170:170] ({(0.325*cos(\t))*cos(-0)-(0.05*sin(\t))*sin(-0)+0.46},{(0.325*cos(\t))*sin(-0)+(0.05*sin(\t))*cos(-0)+0.8});

\draw[black,thick,solid,->] (0.0,0.0) -- (0.0,-1.0);

\draw[black,thick,solid] (0,-1.5) ellipse (0.35 and 0.1);
\draw [black,thick,solid] plot [smooth, tension=0.6] coordinates { (-0.805,-2.8)  (-0.4,-1.9) (-0.35,-1.5)};
\draw [black,thick,solid] plot [smooth, tension=0.6] coordinates { (-0.155,-2.8)  (0.0,-2.65) (0.155,-2.8)};
\draw [black,thick,solid] plot [smooth, tension=0.6] coordinates { (0.805,-2.8)  (0.4,-1.9) (0.35,-1.5)};
\draw[black,thin,dashed] plot[variable=\t,samples=10,domain=-170:170] ({(0.4*cos(\t))*cos(-0)-(0.05*sin(\t))*sin(-0)+0.0},{(0.4*cos(\t))*sin(-0)+(0.05*sin(\t))*cos(-0)-1.9});
\draw[black,thin,dashed] plot[variable=\t,samples=10,domain=-170:170] ({(0.325*cos(\t))*cos(-0)-(0.05*sin(\t))*sin(-0)-0.48},{(0.325*cos(\t))*sin(-0)+(0.05*sin(\t))*cos(-0)-2.8});
\draw[black,thin,dashed] plot[variable=\t,samples=10,domain=-170:170] ({(0.325*cos(\t))*cos(-0)-(0.05*sin(\t))*sin(-0)+0.48},{(0.325*cos(\t))*sin(-0)+(0.05*sin(\t))*cos(-0)-2.8});

\draw[black,thick,solid,->] (0.0,-3.2) -- (0.0,-4.2);

\draw[black,thick,solid] (0,-4.7) ellipse (0.35 and 0.1);
\draw [black,thick,solid] plot [smooth, tension=0.6] coordinates { (0.35,-4.7)  (0.4,-5.1) (0.54,-5.9)};
\draw [black,thick,solid] plot [smooth, tension=0.6] coordinates { (-0.35,-4.7)  (-0.4,-5.1) (-0.54,-5.9)};
\draw[black,thin,dashed] plot[variable=\t,samples=10,domain=-170:170] ({(0.4*cos(\t))*cos(-0)-(0.05*sin(\t))*sin(-0)+0.0},{(0.4*cos(\t))*sin(-0)+(0.05*sin(\t))*cos(-0)-5.1});
\draw[black,thick,solid] (0.54,-5.9) arc (40:-220:0.7);
\draw[black,thick,solid] (0.188,-6.40) arc (20:-200:0.2);
\draw[black,thick,solid] (-0.27,-6.53) arc (-195:-345:0.28);
\end{tikzpicture}
\caption{}
\label{fig:torus}
\end{subfigure}
\caption{Construction of the $(4,0)$ and $(1,1)$ geometries using two and one pairs of pants, respectively. The dashed lines represent horizons of asymptotic regions. Note that each pair of pants is constructed from the process shown in figure \ref{fig:3bdy}, but here the shape of the Riemann surface is shown explicitly.
}
\end{figure}

For instance, to construct the rotating $(4,0)$ geometry, we need two pairs of pants, each having 6 parameters (i.e. the mass and angular momentum of each asymptotic region). We consider the Cauchy slices where both pairs are of the form shown in figure \ref{fig:3bdy}. As shown in figure \ref{fig:4bdy}, if we cut only one asymptotic region in each of the pair of pants and glue the horizons together, this forces the lengths and orientations of the glued horizons to be equal (the $\ell$'s and $\tilde{\ell}$'s of the two glued regions) and introduces two new twist parameters. So, the total number of parameters is 12, which is the correct dimension of the moduli space of the rotating $(4,0)$ geometry. From the resulting Cauchy slice, we can time evolve and obtain the whole required geometry. Similarly, to construct general rotating $(n,0)$ geometries, we need $n-2$ pairs of pants. By cutting $2n-6$ asymptotic regions and gluing them together, we can construct a Cauchy slice of the rotating $(n,0)$ spacetime from which the whole geometry can be obtained by time evolution. One can easily check that the number of parameters in the resulting geometry is the correct dimension of the moduli space, which is $2\left(3n-6\right)$.

In the case of non-zero genus, we consider the simple case of rotating $(1,1)$ spacetime, which was first constructed in \cite{Aminneborg:1998si}. Using a Cauchy slice of a single rotating $(3,0)$ geometry, we can cut two asymptotic regions and then glue their horizons together. The remaining asymptotic region is now the exterior of a rotating BTZ black hole with the topology of a torus behind the horizon, as shown in figure \ref{fig:torus}. One can easily check that this process gives the correct number of dimensions of the moduli space, which is 6 in the case of rotating $(1,1)$ spacetime.

\subsection{Fixed points and the conformal boundary}\label{sec2.2}
We now discuss the action of isometries $\gamma \in \Gamma$ on the conformal boundary of $\text{AdS}_3$, following the method discussed in \cite{Aminneborg:1998si}. Here we will be using the conformal frame
\be
ds_{\text{global}}^2=-dt^2+d\phi^2
\ee
which is naturally related to the global coordinates.

Taking $r\rightarrow \infty$ for a bulk point $p$ \eqref{pembedd} gives a boundary point $p_\partial$. Up to a diverging factor, it is
\begin{align}\label{bdy-p}
p_\partial& \propto \begin{pmatrix}
\cos\phi+\cos t & \sin\phi-\sin t\\
\sin\phi+\sin t & -\cos\phi+\cos t
\end{pmatrix}=2 \begin{pmatrix}
\cos \frac{v}{2}\cos \frac{u}{2} & -\cos \frac{v}{2}\sin \frac{u}{2}\\
\sin \frac{v}{2}\cos \frac{u}{2} & -\sin \frac{v}{2}\sin \frac{u}{2}
\end{pmatrix}=2 \vec{v}\vec{u}^t
\end{align}
where
\be
\vec{v}= \begin{pmatrix}
\cos \frac{v}{2}\\
\sin \frac{v}{2}
\end{pmatrix},\quad
\vec{u}=\begin{pmatrix}
\cos \frac{u}{2}\\
-\sin \frac{u}{2}
\end{pmatrix}
\ee
and $v=t+\phi$ and $u=t-\phi$ are the null coordinates at the boundary. The isometries of interest $\gamma=\left(g_{L},g_{R}\right)\in \Gamma$ are hyperbolic elements with their fixed points at the boundary of $\text{AdS}_3$. Being a fixed point amounts to
\begin{equation}
p_\partial= g_{L}p_\partial g_{R}^t \Rightarrow \vec{v}\vec{u}^t=g_{L} \vec{v}( g_{R}\vec{u})^t,
\end{equation}
where the equality holds up to an overall factor, since we are on the conformal boundary.

This means that we could find fixed points by finding eigenvectors of $g_{L}$ and $g_{R}$. In general, $g_L$ and $g_R$ each have two eigenvectors, and combinations of them give ``corners" of the ``boundary diamond" of $\gamma$ where the action of $\gamma$ takes place. Next, we will illustrate these notions for the BTZ black hole and the three-boundary black hole. Analysis of fixed points for general $(n,g)$ geometries could be performed in a similar manner.

\begin{figure}
\centering
\includegraphics[scale=0.4]{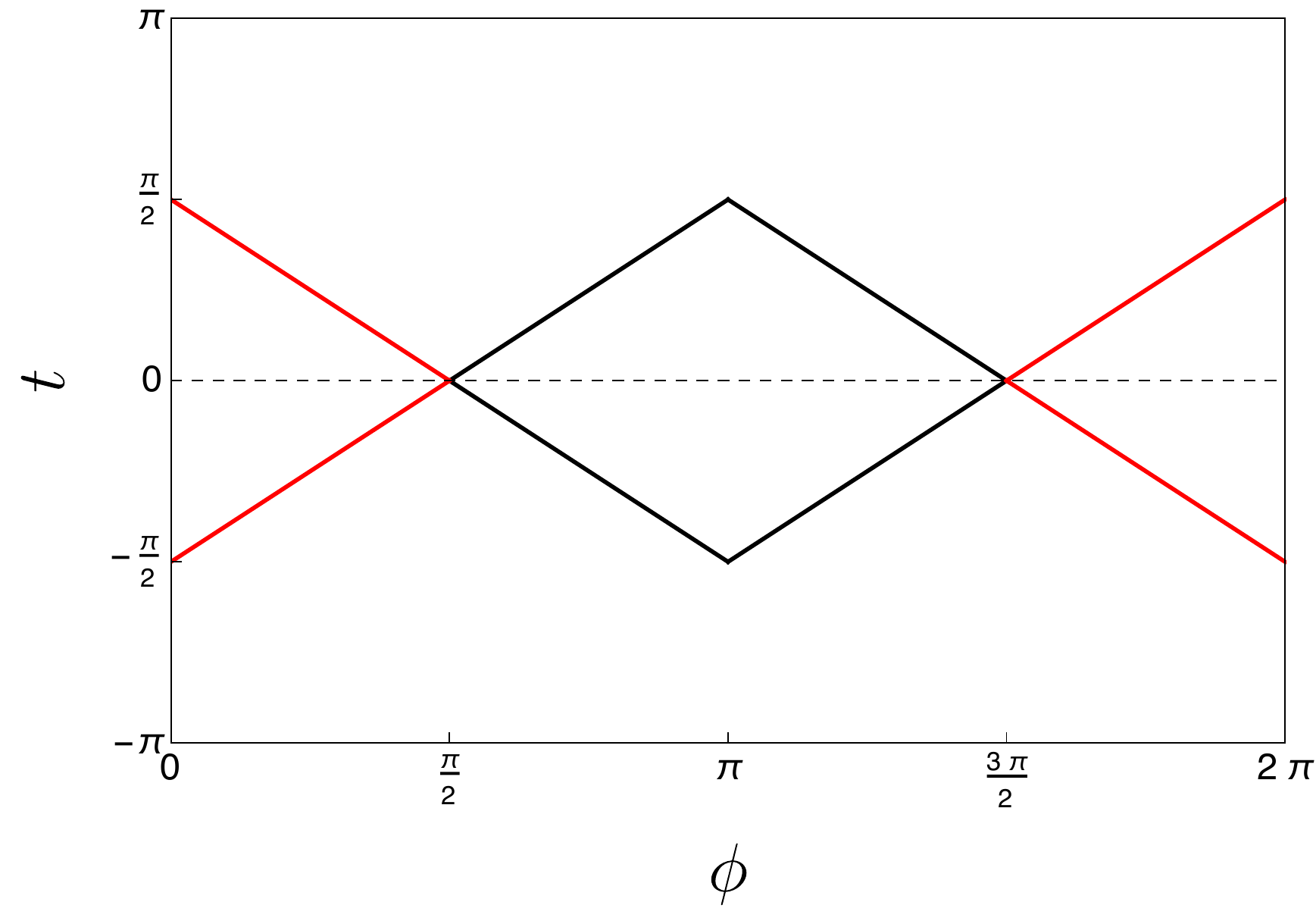}
\caption{Boundary diamonds for the BTZ black hole, where $\phi\sim\phi+2\pi$. As we can see, there are two diamonds, each containing one asymptotic boundary of the fundamental domain.}
\label{BTZdiamonds}
\end{figure}

For the BTZ black hole, all elements of $\Gamma$ are integer powers of $\gamma_{BTZ}$. Both $g_{L,BTZ}$ and $g_{R,BTZ}$ have two eigenvectors
\begin{equation}
g_{L,BTZ}\vec{v}_{\pm}=e^{\pm\ell/2}\vec{v}_{\pm},\quad g_{R,BTZ}\vec{u}_{\pm}=e^{\pm\tilde{\ell}/2}\vec{u}_{\pm}
\end{equation}
where
\begin{equation}
\vec{v}_{\pm}=\frac{1}{\sqrt{2}}\begin{pmatrix}
\pm 1\\
1
\end{pmatrix}, \quad \vec{u}_{\pm}=\frac{1}{\sqrt{2}}\begin{pmatrix}
\pm 1\\
1
\end{pmatrix}.
\end{equation}
As shown in figure \ref{BTZdiamonds}, there only two boundary diamonds for the BTZ black hole, with their left and right corners at $(t=0,\phi=\pi/2)$ and $(t=0,\phi=3\pi/2)$. Inside each diamond, there are infinitely many copies of the fundamental domain, or in other words, the fundamental domain and its images.

For the three-boundary black hole, we could find the fixed points and boundary diamonds in a similar manner. But in this case, we have infinitely many fixed points (and diamonds) since the group $\Gamma$ not only contains elements like $\gamma_i^m,i=1,2$ but also more general ``words" like $\gamma_1^m \gamma_2^n \gamma_1^k...$ etc. 
For $\gamma_i,i=1,2$ we have
\be
g_{iL}.\vec{v}_{\pm,i}=e^{\pm\ell_i/2}\vec{v}_{\pm,i},\quad g_{iR}.\vec{u}_{\pm,i}=e^{\pm\tilde{\ell_i}/2}\vec{u}_{\pm,i}
\ee
with $\vec{v}_{\pm,1}$ and $\vec{u}_{\pm,1}$ the same as those of the BTZ black hole, and
\begin{equation}
\label{H2Diamonds}
\vec{v}_{\pm,2}=\frac{1}{\sqrt{1+e^{2\alpha}}}\begin{pmatrix}
\pm e^{\alpha}\\
1
\end{pmatrix}, \quad \vec{u}_{\pm,2}=\frac{1}{\sqrt{1+e^{2\tilde{\alpha}}}}\begin{pmatrix}
\pm e^{\tilde{\alpha}}\\
1
\end{pmatrix}.
\end{equation}

\begin{figure}[t]\centering
\begin{subfigure}{0.49\textwidth}\centering
\includegraphics[width=0.85\linewidth]{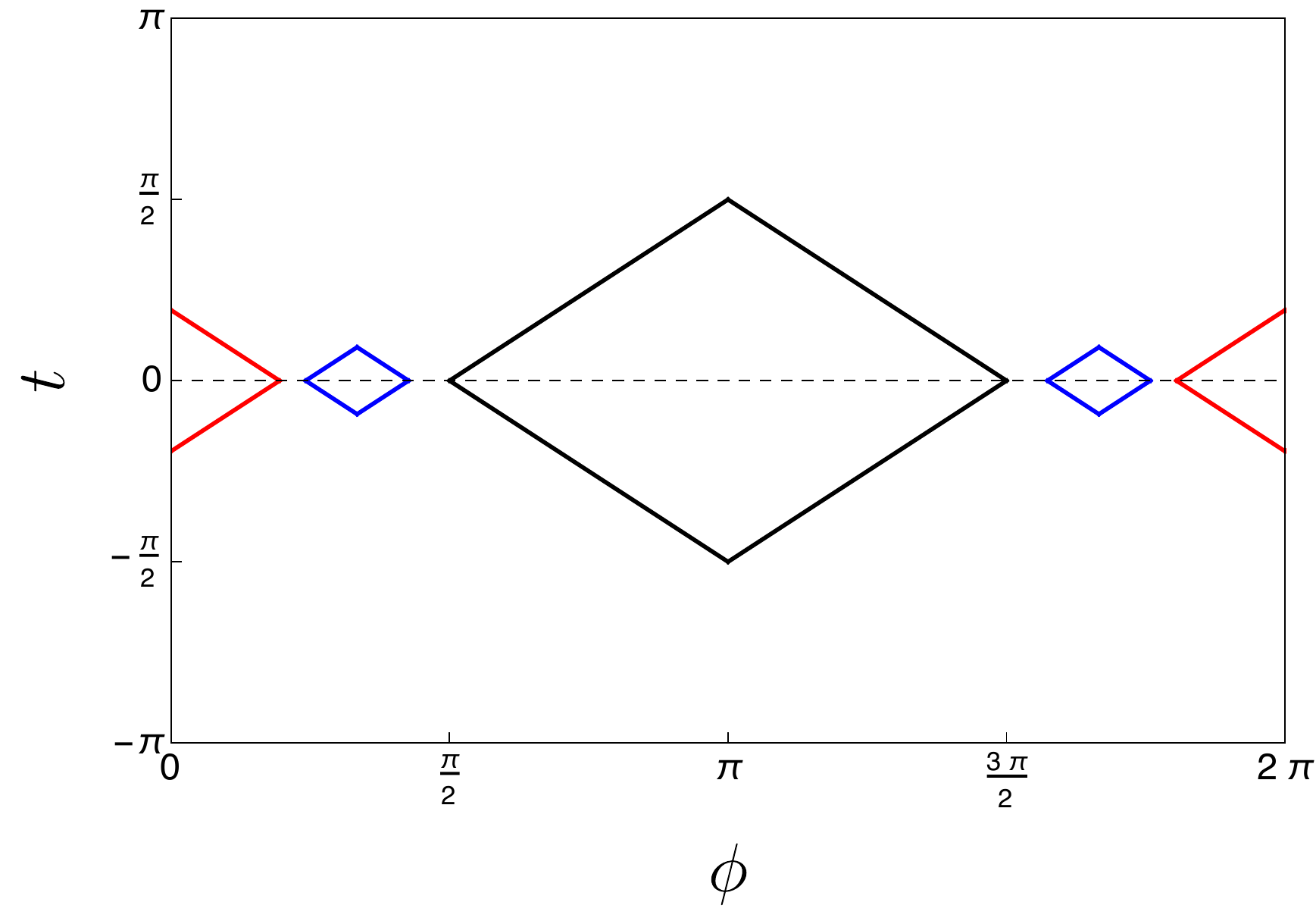}
 \caption{}
\label{fig:diamondsNonRot}
\end{subfigure}
\begin{subfigure}{0.49\textwidth}\centering
\includegraphics[width=0.85\linewidth]{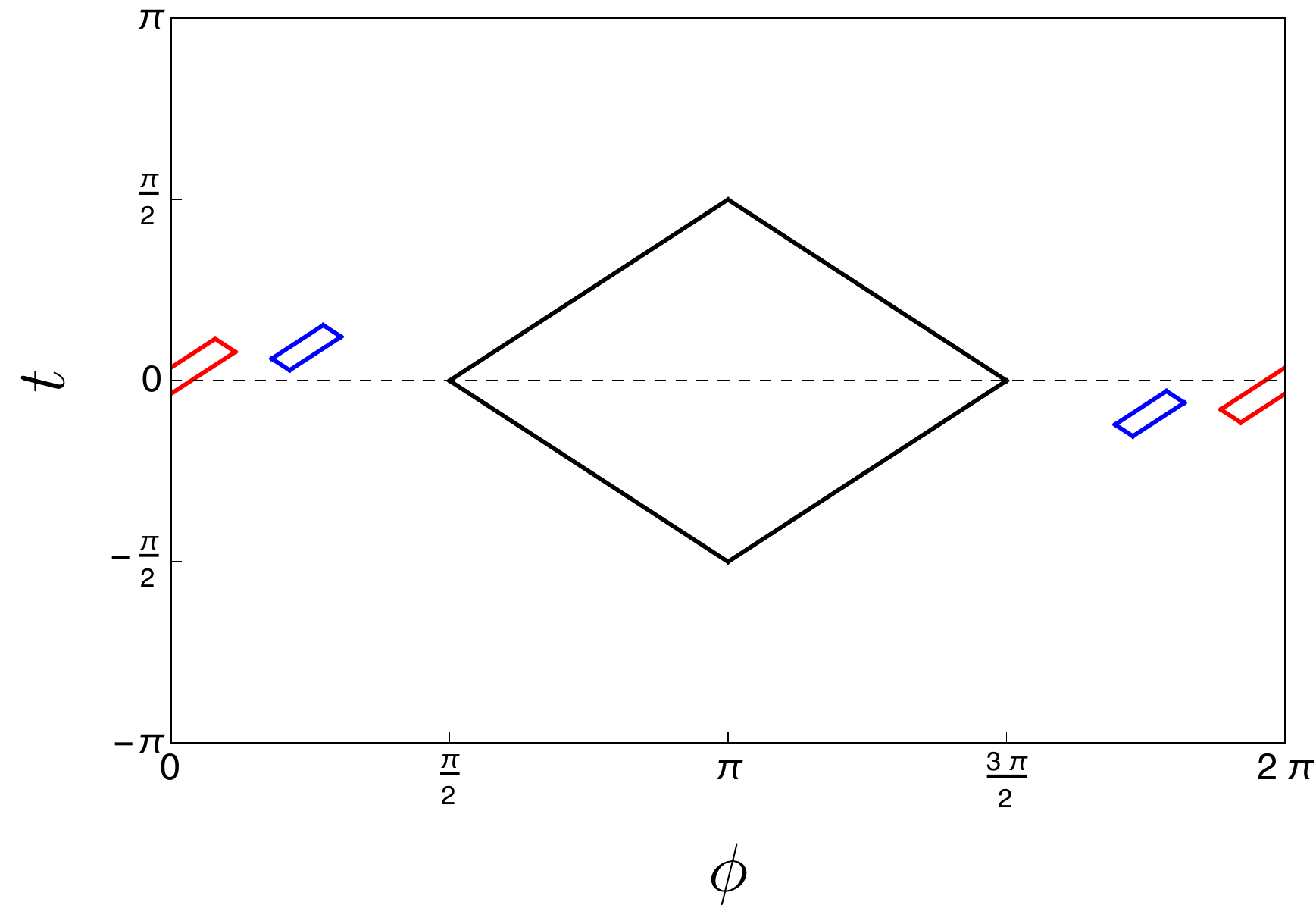}
\caption{}
\label{fig:diamondsRot}
\end{subfigure}
\caption{The fundamental diamonds of $(3,0)$ geometry at the boundary of AdS$_3$ in global coordinates. The fixed points $p_{++,i}$, $p_{--,i}$, $p_{-+,i}$, and $p_{+-,i}$ correspond to the corners of the diamonds. The diamonds of regions 1, 2, and 3 are bounded by black, red, and blue lines respectively. In (a), the parameters are $\ell_i=\tilde{\ell_i}=3$ for the non-rotating case, and in (b) the parameters are $\ell_i=3\tilde{\ell_i}=3$ for the rotating case.  
}\label{fig:diamonds}
\end{figure}
For the three-boundary black hole, the three asymptotic boundaries of the fundamental domain are contained in the diamonds which we call ``fundamental diamonds" generated by $\gamma_i,i=1,2,3$. Other diamonds will be dubbed ``image diamonds". In figure \ref{fig:diamonds}, we show the fundamental diamonds of the three-boundary black hole. The corners of the fundamental diamonds can be found from
\be\label{pFixed}
p_{++,i}=\vec{v}_{+,i}\vec{u}_{+,i}^t\quad,\quad p_{+-,i}=\vec{v}_{+,i}\vec{u}_{-,i}^t\quad,\quad p_{-+,i}=\vec{v}_{-,i}\vec{u}_{+,i}^t\quad,\quad p_{--,i}=\vec{v}_{-,i}\vec{u}_{-,i}^t.
\ee
where again $i=1,2,3$.

For any point $p_\partial$ on  the $i^\text{th}$ asymptotic region of the fundamental domain, there are two types of image points under the group action:

\begin{enumerate}
\item Points that are in the same fundamental diamond as $p_\partial$: these points are generated by acting on $p_\partial$ with isometries that only involve integer powers of $\gamma_i$;
\item Points that are in the image diamonds: these points are generated by acting with other kinds of isometries on $p_\partial$.
\end{enumerate}


Although it is hard to find the explicit locations of all of the image diamonds, they must all lie between diamonds 1 and 2, and topological censorship guarantees that any pair of diamonds must be spacelike separated. The boundary distance from the left corner of diamond 1 $(p_{++,1})$ to the right corner of diamond 2 $(p_{++,2})$ is
\begin{equation}
\begin{aligned}
d_{\text{bdy}}(p_{++,1},p_{++,2})&= \sqrt{\left|(u_{++,1}-u_{++,2}) (v_{++,1}-v_{++,2})\right|}\\
&=\sqrt{\left(\frac{\pi}{2}-2\tan^{-1} e^{-\alpha}\right)\left(\frac{\pi}{2}-2\tan^{-1} e^{-\tilde{\alpha}}\right)}.
\end{aligned}
\end{equation}
When $\alpha$ and $\tilde{\alpha}$ are small (i.e. $\ell_i$ and $\tilde{\ell}_i$ are large), to leading order, the distance is
\begin{equation}\label{dbdyalphasqrt}
d_{\text{bdy}}(p_{++,1},p_{++,2})= (\alpha \tilde{\alpha})^{\frac{1}{2}}+\mathcal{O}((\alpha \tilde{\alpha})^{\frac{3}{2}}).
\end{equation}


Given a choice of the boundary conformal frame, we can also define the regularized geodesic distance through the bulk between boundary points. 
First, note that for any $2\times 2$ matrix $p$ with $\det p =1$ we have
\be
p^{-1}=R_\perp p^t R_\perp^t,\quad\text{where}\quad R_\perp=\begin{pmatrix}
0& -1\\
1& 0
\end{pmatrix}
\ee
Also, the elements of a matrix $p$ of any bulk point scales linearly with $r$. So, in the limit $r\rightarrow \infty$ we find
\begin{align}
d_{\text{bulk}}(p_1, p_2)&=\cosh ^{-1}\left(\frac{\operatorname{Tr}\left(p_1^{-1} p_2\right)}{2}\right)\nonumber\\
&=\cosh ^{-1}\left(\frac{\Tr\left(R_\perp p_1^t R_\perp^t p_2\right)}{2}\right)\nonumber\\
&= \log\left(r^2\right)+\log\left(\Tr\left(R_\perp p_{\partial 1}^t R_\perp^t p_{\partial 2}\right)\right) +\mathcal{O}\left(r^{-2}\right)\nonumber\\
&= \log\left(r^2\right)+\log\left(4\Tr\left(R_\perp {\left(\vec{v}_1\vec{u}_1^t\right)}^t R_\perp^t \left(\vec{v}_2\vec{u}_2^t\right)\right)\right) +\mathcal{O}\left(r^{-2}\right)
\end{align}
To find the renormalized boundary geodesic distance, we subtract $\log\left(r^2\right)$ then take the $r\rightarrow\infty$ limit, giving
\begin{align}\label{lreg}
d_{\text{ren}}^{\text{global}}(p_{1\partial},p_{2\partial})&=\log\left(4\left(\vec{u_1}^\perp . \vec{u_2}\right)\left(\vec{v_1}^\perp . \vec{v_2}\right)\right),
\end{align}
where
\be
\vec{u}^\perp=R_\perp \vec{u}\quad \text{and}\quad \vec{v}^\perp=R_\perp\vec{v}.
\ee
Similarly, the renormalized geodesic distance between a bulk point $p$ and a boundary point $q_\partial=2\ \vec{v}\vec{u}^t$ is given by
\be\label{lreg2}
d_{\text{ren}}^{\text{global}}(p,q_\partial)=\log\left(\operatorname{Tr}\left(p^{-1} q_\partial \right)\right) =\log\left(2\operatorname{Tr}\left(p^{-1} \vec{v}\vec{u}^t \right)\right).
\ee
An important question is finding the corresponding expressions to the renormalized geodesic distances \eqref{lreg}-\eqref{lreg2} for the boundary of an asymptotic region that is in the BTZ conformal frame $ds_{BTZ}^2=-dt_B^2+d\phi_B^2$. This question is resolved in subsection \ref{sec:BTZdistance}.

\subsection{Geodesic distances in the BTZ conformal frame}
\label{sec:BTZdistance}
In this subsection, we calculate the renormalized geodesic distance from a bulk point $p$ to a boundary point $q_\partial$ that is in the BTZ conformal frame. We assume that $q_\partial$ is on the boundary of the fundamental domain,  so it is in one of those fundamental diamonds defined in section \ref{sec2.2}. In that diamond, we choose the BTZ conformal frame, and the renormalized distance we calculate here is compatible with that frame. We also assume that $p$ and $q_\partial$ are spacelike separated so that we use \eqref{Lspacelike} rather than \eqref{Ltimelike} to calculate the distance.

First let us work out the conformal transformation between the AdS global conformal frame and the BTZ frame. For simplicity, we first study a boundary diamond of the BTZ black hole, as shown in figure \ref{BTZdiamonds}. Then we convert our results to smaller diamonds using isometries.

Recall that global AdS$_3$ and the BTZ coordinates are related to the embedding coordinates via \eqref{globalAdS3} and \eqref{BTZembed}. On the boundary where both radial coordinates go to infinity one finds
\begin{equation}
Y/X=\tan \phi =\frac{\sinh\frac{-\tilde{\ell} u_B+\ell v_B}{4\pi}}{\cosh\frac{\tilde{\ell} u_B+\ell v_B}{4\pi}},\quad V/U=\tan t =\frac{\sinh\frac{\tilde{\ell} u_B+\ell v_B}{4\pi}}{\cosh\frac{-\tilde{\ell} u_B+\ell v_B}{4\pi}},
\end{equation}
where $u_B=t_B-\phi_B$, $v_B=t_B+\phi_B$. Then, using null coordinates $u=t-\phi$ and $v=t+\phi$ on the global AdS$_3$ boundary, the above equations simplify to
\begin{equation}
u=\tan^{-1} \sinh \frac{\tilde{\ell} u_B}{2\pi}, \quad v=\tan^{-1} \sinh \frac{\ell v_B}{2\pi}.
\end{equation}
These observations allow us to compute the conformal transformation between the two conformal frames,
\begin{equation}
ds^2_{\text{global}}=-dudv=\Omega^2(- du_B dv_B)=\Omega_u^2 \Omega_v^2(- du_B dv_B)=\Omega_u^2 \Omega_v^2ds^2_{\text{BTZ}}
\end{equation}
where the conformal factor $\Omega^2$ factorizes into the ``left-moving" and ``right-moving" conformal factors
\begin{equation}
\Omega^2_u=\frac{\tilde{\ell}}{2\pi \cosh \frac{\tilde{\ell} u_B}{2\pi}}=\frac{\tilde{\ell}}{2\pi} \cos u ,\quad \Omega^2_v=\frac{\ell}{2\pi \cosh \frac{\ell v_B}{2\pi}}=\frac{\ell}{2\pi} \cos v.
\end{equation}

As we can see, when $u=\pm \frac{\pi}{2}$ or $v=\pm\frac{\pi}{2}$ either $u_B$ or $v_B$ will diverge and the conformal factors vanish. This marks the boundary of the ``boundary diamond" being considered.  Note also that the conformal factors reach their maximal value at the ``center"of the diamond where $u=0$ and $v=0$.

For any wormhole, each asymptotic region is isometric to the exterior of some BTZ solution.  So up to conformal transformations each boundary of any wormhole is identical to the boundary diamonds just described. While this always yields another diamond, the ranges $\Delta u$ and $\Delta v$ for general boundary diamonds can differ from $\pi$.  But we can use the appropriate conformal transformations to generalize the analysis above.

Indeed, for the construction described in section \ref{sec:AdS3}, the relevant conformal transformations are those induced by isometries of AdS$_3$.  Recall that the  generators of AdS$_3$ isometries act on the boundary as
\begin{equation}
2J_{1} =-\left(J_{X U}-J_{Y V}\right)=\sin v \partial_{v}\equiv \partial_x, \quad 2\tilde{J}_{1} =-\left(J_{X U}+J_{Y V}\right)=\sin u \partial_{u}\equiv \partial_y,
\end{equation}
where we have defined
\begin{equation}
x=\log \tan \frac{v}{2},\quad y=\log \tan \frac{u}{2}.
\end{equation}
These actions, written here as translations in $x$ and $y$, change the size of the boundary diamond. We analyze this in detail for  $v$ direction below, from which corresponding expressions for the $u$ direction follow from the symmetry $u \leftrightarrow v$.

We first note that translating $x$ by $x_0=\log \tan \frac{v_0}{2}$ changes the diamond boundaries from $v=\pm \frac{\pi}{2}$ to $v=\pm v_0$. Denoting the left-moving coordinate in the new diamond by $v'$ we have
\begin{equation}
\tan \frac{v'}{2}=\tan \frac{v}{2}\tan \frac{v_0}{2}.
\end{equation}
Here we assume $v_0<\frac{\pi}{2}$ and $v'=\pm v_0=\pm \frac{\Delta v}{2}$ are the boundaries of the new diamond given by the images of $v=\pm \frac{\pi}{2}$. This relation implies
\begin{equation}
dv'=\frac{1-\cos v' \cos v_0}{\sin v_0}dv.
\end{equation}
The left-moving conformal factor then becomes
\begin{equation}
\Omega_v^2=\left( \frac{\ell}{2 \pi} \cos v\right) \left(\frac{1-\cos v' \cos v_0}{\sin v_0}\right) =\frac{\ell}{2\pi} \frac{\cos v'-\cos v_0}{\sin v_0}
\end{equation}
Inside a diamond, it is bounded by
\begin{equation}
\Omega_v^2\leq \frac{\ell}{2 \pi} \tan\frac{v_0}{2}=\frac{\ell}{2 \pi} \tan\frac{\Delta v}{4},
\end{equation}
where the equality holds at $v'=0$. When a diamond has a small size, this bound is approximately
\begin{equation}
\Omega_v^2\lesssim \frac{\ell v_0}{4 \pi}=\frac{\ell \Delta v}{8 \pi}.
\end{equation}
Also inside a diamond, when the point is close to one edge of the diamond (i.e. when $v'$ is close to $v_{\text{bdy}}=v_0$ or $-v_0$), $\Omega_v^2$ has the expansion
\begin{equation}\label{Omegabdy}
\Omega_v^2=\frac{\ell}{2\pi}(|v'-v_{\text{bdy}}|)+ \mathcal{O}((v'-v_{\text{bdy}})^2).
\end{equation}

Similar relations hold for the $u$ direction. Diamonds that are not centred at $v=0,u=0$ can of course be translated to this standard position using the boundary isometries $\partial_v$ and $\partial_u$ so that corresponding bounds and expressions apply.

As discussed in section \ref{sec2.2}, if we regulate a boundary point $q_{\partial}$ by moving it to a finite global AdS$_3$ radial coordinate $r$, the geodesic distance between a bulk point $p$ and a boundary point $q_{\partial}$ is
\begin{equation}
\begin{aligned}
d_{\text{bulk}}(p, q) &=\cosh ^{-1}\left(\frac{\operatorname{Tr}\left(p^{-1} q\right)}{2}\right) \\
&=\log (r)+\log \left(\operatorname{Tr}\left(p^{-1} q_{\partial}\right)\right)+\mathcal{O}\left(r^{-2}\right).
\end{aligned}
\end{equation}

To renormalize the distance in the BTZ conformal frame associated with a given asymptotic region of our wormhole, we should take the limit $r\rightarrow \infty$ after subtracting  $\log r_B$ from the above expression for a properly chosen radial coordinate $r_B$ associated to the boundary diamond containing $q_\partial$.

In Fefferman-Graham coordinates, when we transform between the global and BTZ conformal frames, to leading order in $z$, we have $z_B=z/|\Omega|$. Also, to leading order, $z \sim 1/r$ and $z_B \sim 1/r_B$, so we have $r_B\sim r |\Omega|=r |\Omega_u\Omega_v|$. A properly defined renormalized geodesic distance is thus given by
\begin{equation}
\label{drenBTZ}
d_{\text{ren}}^{\text{BTZ}}(p,q_\partial)=\log \left(\operatorname{Tr}\left(p^{-1} q_{\partial}\right)\right)-\log|\Omega_u\Omega_v|=d_{\text{ren}}^{\text{global}}(p,q_\partial)-\log|\Omega_u\Omega_v|.
\end{equation}

\subsection{The hot limit of multi-boundary wormholes}\label{sec2.3}
In order to construct multi-boundary traversable wormholes in section \ref{sec:MultiTrav}, we will need to take  a limit that produces the following features: 1) two horizons are separated only by an exponentially thin causal shadow over a sufficiently large region of those horizons, and 2) we can find a point $q_\partial$ on the boundary of the fundamental domain such that the conformal factors $\Omega^2=\Omega_u^2 \Omega_v^2$ associated with its non-trivial images under the group $\Gamma$ are exponentially small. For reasons that will be clear below, we use the term ``hot limit'' to describe this limit for any $(n,g)$.

For multi-boundary wormholes with trivial topologies, we choose to take a limit where all $\ell_i$ and $\tilde{\ell}_i$ are large, with $\ell_i/\tilde{\ell}_i$ fixed (i.e. $M_i/J_i$ fixed)\footnote{For wormholes with internal parameters (i.e. non-trivial topologies or with $n>3$), the proper limit will also involve taking certain internal parameters to be large, in addition to having $\ell_i$ and $\tilde{\ell}_i$ large, with $\ell_i/\tilde{\ell}_i$ fixed. We will discuss this briefly in section \ref{sec:discussion}.}. 
 In the case without rotation, this is exactly the ``hot limit" considered in \cite{Marolf:2015vma}. In the case with rotation, this is also a limit where the temperatures in all asymptotic regions are large. It also implies that all horizon lengths are large compared to the AdS scale (although the converse is not necessarily true). We explain the two advertised features below, using the three-boundary wormhole as our main example.

First, we study the minimal distance between two neighbouring horizons.  For non-rotating $(3,0)$ geometries, this has been computed in \cite{Marolf:2015vma} by focusing on the half-plane of the $t=0$ slice. The minimal distance $d_{ij}$ between horizons $H_i$ and $H_j$ depends on the horizon lengths, and is given by
\be\label{dij}
\cosh d_{ij}=\frac{\cosh\left(h_i/2\right)\cosh\left(h_j/2\right)+\cosh\left(h_k/2\right)}{\sinh\left(h_i/2\right)\sinh\left(h_j/2\right)}.
\ee
Applying \eqref{dij} to horizons $H_1$ and $H_2$ in our construction, we have from \eqref{alphaEq} that
\begin{equation}\label{d12alphaEq}
d_{12}=\alpha=\tilde{\alpha}.
\end{equation}
In appendix \ref{appdij}, we generalize \eqref{dij} to the case with rotations, where the minimal distance between horizons $H_1$ and $H_2$ was shown to be given simply by
\begin{equation}
d_{12}=\frac{\alpha+\tilde{\alpha}}{2}.
\end{equation}
Other minimal horizon distances can be found from this expression by simple permutations. It can be easily shown that $\alpha$ and $\tilde{\alpha}$ are exponentially small in the hot limit, and that $d_{ij}$ is as well. As a special case, when all $\ell_i=\ell$ and $\tilde{\ell}_i=\tilde{\ell}$ are large, we have $\alpha\sim 2e^{-\ell/4}, \tilde{\alpha}\sim 2e^{-\tilde{\ell}/4}$ and $d_{ij}\sim e^{-\ell/4}+e^{-\tilde{\ell}/4}$. Furthermore, in this limit, it was found \cite{Marolf:2015vma} that the distance between the horizons is exponentially small over a large subset $D_\phi$ of the angular domain, for which the lateral extent along each horizon is large compared with the AdS scale. In appendix \ref{appdij}, we show that this feature also applies in the rotating case. In addition, we show there that this is no longer the case when only one of $\ell_i$ or $\tilde{\ell}_i$ are taken to be large.  The latter limit makes the horizons large but the horizon temperatures remain bounded\footnote{This has some interesting consequences for the extremal limit that we briefly discuss in section \ref{sec:discussion}.}.
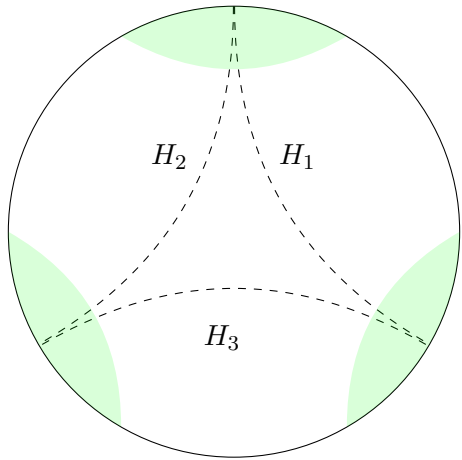
\begin{figure}
\centering
\begin{tikzpicture}
\draw (0,0) circle[radius=3];
\draw [dashed] (-0.0075,3) to[out=270,in=30] (-2.6,-1.5);
\draw [dashed] (0.0075,3) to[out=270,in=150] (2.6,-1.5);
\draw [dashed] (-2.55,-1.5) to[out=30,in=150] (2.55,-1.5);
\node[text width=0.8cm] at (1,1) {$H_1$};
\node[text width=0.8cm] at (-.7,1) {$H_2$};
\node[text width=0.8cm] at (0,-1.41) {$H_3$};
\fill [fill=green, opacity=0.15] (1.5,-2.6) to [out=90,in=210] (3,0) arc (0:-60:3);
\fill [fill=green, opacity=0.15] (-1.5,-2.6) to[out=90,in=-30] (-3,0) arc (180:240:3);
\fill [fill=green, opacity=0.15] (-1.5,2.6) to[out=-30,in=210] (1.5,2.6) arc (60:120:3);
\end{tikzpicture}
\caption{A schematic diagram of the $t=0$ slice of a three-boundary black hole in the hot limit. For any pair of horizons (dashed lines), there is a large region which we call $D_\phi$ (highlighted in green) where the horizons are exponentially close to each other. The causal shadow is the region bounded by the three horizons.}
\end{figure}

Similar results also hold in the case of a general $n$-boundary black hole. As discussed in section \ref{sec2.1}, a general $(n,0)$ spacetime with $n\geq 3$ can be constructed from $n-2$ copies of $(3,0)$ geometry. Here we compute the minimal distance $d_{ij}$ between any two horizons $H_i$ and $H_j$ that live in a single copy of $(3,0)$ geometry, though we comment on the more general case below. For $n>3$ the third horizon $H_k$ in this copy will become part of the causal shadow of the new $(n,0)$ geometry and its length $h_k$ will be one of the parameters of the moduli space associated with the casual shadow region. Therefore, the same minimal distance $d_{ij}$ between horizons $H_i$ and $H_j$ as in the $(3,0)$ geometry will hold. Choosing $h_k\ll h_i+h_j$ as in the hot limit above, $d_{ij}$ will again be exponentially small.  In the more general case\footnote{We have not yet discussed the case $g>0$ in detail, but see section \ref{sec:discussion} for comments.} $g\neq 0$,  or for two horizons in the $(n,0)$ geometry which are separated by an intervening extremal surface\footnote{In the case without time-symmetry, this means that the intervening extremal surface lies in the domain of dependence of any partial (connected) Cauchy slice $\Sigma$ for which $\partial \Sigma = H_i \cup H_j$. } and thus which lie in distinct copies of the $(3,0)$ geometry, taking the hot limit for each copy of the $(3,0)$ geometry allows us to write the separation between $H_i$ and $H_j$ as the union of a fixed finite number of exponentially small separations.  Thus we find the separation between $H_i$ and $H_j$ to be exponentially small in the hot limit for all $n,g$.

The other important feature of the geometry in the hot limit is that we can find points $q_\partial$ on the boundary for which the non-trivial image points $q_\partial^{\text{image}}$ all have conformal factors that are exponentially small.  This property will be established below, but its important consequence follows from equation \eqref{drenBTZ} governing the renormalized distance between $p$ and $q_\partial^{\text{image}}$ in BTZ frame.  From this it follows that
\be\label{blahblack}
d_{\text{ren}}^{\text{BTZ}}(p,q_\partial^{\text{image}})=\log \left(\operatorname{Tr}\left(p^{-1} q_{\partial}^{\text{image}}\right)\right)-\log|\Omega_u\Omega_v|=d_{\text{ren}}^{\text{global}}(p,q_\partial^{\text{image}})-\log|\Omega_u\Omega_v|.
\ee
Here $\Omega_u$ and $\Omega_v$ are the conformal factors associated with $q_\partial^{\text{image}}$. So when we have a bulk point $p$ that is in the same asymptotic region as $q_\partial$, in the BTZ frame, the exponentially small conformal factors associated with the images require $d_{\text{ren}}^{\text{BTZ}}(p,q_\partial^{\text{image}}) > d_{\text{ren}}^{\text{BTZ}}(p,q_\partial)$  with their difference being linear in $\ell_i$ and $\tilde{\ell}_i$.

To show for appropriate $q_\partial$ that the conformal factor associated with non-trivial images is exponentially small in the hot limit, recall from section \ref{sec2.2} that the image points are classified into two types. We will take $q_\partial$ to lie in the fundamental domain (for which the boundary diamond is not small).  We first treat image points that lie in other boundary diamonds (i.e. image diamonds).
Recall from section \ref{sec:BTZdistance} that the associated conformal factors satisfy
\be\label{conformalBound}
\Omega_u^2\leq \frac{\tilde{\ell} \Delta u^{\text{image}}}{8 \pi} \quad\text{and}\quad \Omega_v^2\leq \frac{\ell \Delta v^{\text{image}}}{8 \pi}
\ee
where $\Delta u^{\text{image}}$ and $\Delta v^{\text{image}}$ determine the size of the diamond to which $q_\partial^{(m)}$ belongs.
Note that since $d_{\text{bdy}}(p_{++}^{\text{image}},p_{--}^{\text{image}})=\sqrt{\Delta u^{\text{image}}\Delta v^{\text{image}}},$ equation \eqref{conformalBound} implies that
$\Omega_u \Omega_v \le \frac{\sqrt{\tilde{\ell}\ell} }{8 \pi} d_{\text{bdy}}(p_{++}^{\text{image}},p_{--}^{\text{image}}).$


Let us take the $(3,0)$ geometry as our example.  There all the image diamonds lie between diamonds 1 and 2 and are spacelike separated from them. Then, using \eqref{dbdyalphasqrt}, we have in the hot limit
\be
d_{\text{bdy}}(p_{++}^{\text{image}},p_{--}^{\text{image}})<d_{\text{bdy}}(p_{++,1},p_{++,2})\sim\sqrt{\alpha \tilde{\alpha}}.
\ee
Therefore
\begin{equation}
\Omega^2=\Omega_u^2 \Omega_v^2  \lesssim \frac{\tilde{\ell}\ell}{64\pi^2}\alpha \tilde{\alpha}.
\end{equation}
In the hot limit, $\Omega^2$ is exponentially small. As a special case, when $\ell_i=\ell$ and $\tilde{\ell}_i=\tilde{\ell}$ we have $\Omega^2\lesssim e^{-(\ell+\tilde{\ell})/4}$, and since $d_{\text{ren}}^{\text{global}}=\mathcal{O}(1)$ we also have $d_{\text{ren}}^{\text{BTZ}}\gtrsim \mathcal{\ell+\tilde{\ell}}$.

%

The remaining case to consider occurs
when $q_\partial^{\text{image}}$ belongs to the same boundary diamond as $q_\partial$. Let us take $q_\partial$ to lie at some fixed boundary location independent of $\ell_i, \tilde \ell_i$. Then in the hot limit the analysis of section \ref{sec:BTZdistance}  requires $q_\partial^{\text{image}}$ to be exponentially close to one of the fixed points associated with the corners of the fundamental diamond.  Recall from \eqref{Omegabdy} that when this is the case the conformal factors can be approximated as
\begin{equation}\label{OmegaOmegauv}
\Omega_u^2\simeq \frac{\tilde{\ell}}{2 \pi}\left(\left|u-u_{\mathrm{bdy}}\right|\right)\quad \text{and/or}\quad \Omega_{v}^{2}\simeq \frac{\ell}{2 \pi}\left(\left|v-v_{\mathrm{bdy}}\right|\right),
\end{equation}
where $u_{\mathrm{bdy}},v_{\mathrm{bdy}}$ are the coordinates of the relevant corner.

We will show that these conformal factors will be exponentially small and that the renormalized distance to $q_\partial^{\text{image}}$ will be large. In the $(3,0)$ geometry we may derive an explicit expression by recalling the action of the quotient construction on boundary diamonds.  In particular, the quotient of any such diamond is a cylinder.  We may thus discuss a `fundamental domain' within the boundary diamond which we take to be an open set that covers the cylinder precisely once (or, at least, up to a set of measure zero associated with the boundary of the fundamental domain).  We will also choose this domain to be centered at the origin $u,v=0$ and to have a simple form.

The details of such a fundamental domain were computed in   \cite{Balasubramanian:2014hda} for the case where the bulk is a non-rotating BTZ black hole. On the $t=0$ slice, a corresponding fundamental domain in the bulk may be taken to lie between the codimension-1 surfaces
\begin{equation}
\label{eq:fdedges}\phi= \pi \pm \sin^{-1}\left(\tanh\left(\pi r_+\right)\right).
\end{equation}
As a result, the maximal boundary distance $d_\partial$ between the boundary limit of \eqref{eq:fdedges} and the left/right corner of the diamond is
\be\label{dpartialnonrot}
d_\partial=\cos^{-1}\left(\tanh\left(\pi r_+\right)\right).
\ee
In the case of rotation, one can show that this expression generalizes to\footnote{The idea is to realize that, since $\gamma_{BTZ}$ defined in \eqref{BTZgamma} maps the two boundaries of the fundamental domain to each other, then $\gamma_{BTZ}^{1/2}$ will map the boundary centre of the fundamental domain to one of the boundary corners of the fundamental domain. This centre point, in global coordinates, is $(t=0,\phi=\pi)$. Acting on this point with $\gamma_{BTZ}^{1/2}$ gives the coordinates of the corner of the fundamental domain at the boundary, from which we calculate $d_\partial$.}
\be\label{dpartialrot}
d_\partial=\sqrt{\left(\cos^{-1}\tanh \frac{\ell}{2}\right) \left(\cos^{-1}\tanh \frac{\tilde{\ell}}{2}\right)}
\ee
 Note that this equation reduces to \eqref{dpartialnonrot} when $\ell=\tilde{\ell}$, using \eqref{horizonLength}. In the hot limit we find $d_\partial\sim 2e^{-(\ell+\tilde{\ell})/4}$.  Since every domain of outer communication (i.e., every region outside the black hole) is isometric to the domain of outer communication for some BTZ black hole, the corresponding expressions will also hold for our multi-boundary wormholes.


Without loss of generality, we assume that $\ell_1\leq\ell_2,\ell_3$ and $\tilde{\ell}_1\leq\tilde{\ell}_2,\tilde{\ell}_3$. So, from \eqref{dpartialrot}, the largest $d_\partial$ will occur for diamond 1, where it is given by \eqref{dpartialrot} with $\ell$ and $\tilde{\ell}$ replaced by $\ell_1$ and $\tilde{\ell}_1$, respectively. In particular, if $\epsilon$ is the distance between $q_\partial^{(m)}$ and the fixed point of the fundamental diamond, then $\epsilon<d_\partial$. Furthermore, from \eqref{OmegaOmegauv}, we have $\Omega^2\sim\epsilon^2$. This provides a lower bound on $d_{\text{ren}}^{\text{BTZ}}(p,q_\partial^{\text{image}})$ that in the hot limit yields
\be
d_{\text{ren}}^{\text{BTZ}}(p,q_\partial^{\text{image}})\geq -\log\Omega^2\sim -\log\epsilon^2 \geq -\log d_\partial^2 \gtrsim \ell_1+\tilde{\ell}_1
\ee
This verifies explicitly that the conformal factors associated with $q_\partial^{\text{image}}$ are exponentially small in the hot limit , whether $q_\partial^{\text{image}}$ is in an image diamond or in the fundamental diamond. As a consequence, $d_{\text{ren}}^{\text{BTZ}}(p,q_\partial^{\text{image}})\gtrsim\ell+\tilde{\ell}$.

\subsection{The CFT dual of $(n,g)$ geometries}\label{sec2.4}
The bulk $(n,g)$ spacetime is dual to a CFT state $\ket{\Sigma_{n,g}}\in\mathcal{H}_1\otimes\dots\otimes\mathcal{H}_n$, where $\mathcal{H}_i$ is the Hilbert space of a CFT state on a circle. In the energy eigenbasis, this state can be expressed as\footnote{Note that, for simplicity of notation, we are ignoring rotation for a moment. However, these equations can easily be generalized to the case of rotation.}
\be\label{bdyCFT}
\ket{\Sigma_{n,g}}=\sum_{i_1,\dots,i_n}A_{i_1,\dots,i_n}\ket{i_1}_1\dots\ket{i_n}_n
\ee
where the coefficient $A_{i_1,\dots,i_n}$ is a function of the $2(6g-6+3n)$ moduli of rotating $(n,g)$ geometry. A Cauchy slice of $(n,g)$ spacetime is a Riemann surface $\Sigma_{n,g}$ with $n$ boundaries and genus $g$. Suppose that the state of the CFTs at the $n$ boundaries is $\ket{\phi_1\dots\phi_n}\in\mathcal{H}_1\otimes\dots\otimes\mathcal{H}_n$. In the large temperature limit, the gravitational path integral over the Euclidean Riemann surface with boundary conditions fixed by $\ket{\phi_1\dots\phi_n}$ is dominated by the fully-connected bulk geometry, which by Wick rotation gives a Cauchy slice $\Sigma_{n,g}$ that can give the full $(n,g)$ spacetime by Lorentzian time-evolution - see \cite{Balasubramanian:2014hda,Marolf:2015vma,Maxfield:2016mwh} for details. Varying the moduli changes the dominant bulk geometry in the gravitational path integral, which induces first-order phase transitions that generalize the Hawking-Page transition \cite{Hawking:1982dh} in the $(2,0)$ spacetime. For example, for sufficiently large temperatures, the CFT state dual to the BTZ black hole is a thermofield-double state and \eqref{bdyCFT} becomes \cite{Maldacena:2001kr}
\be
\ket{\Sigma_{2,0}}=\sum_{i}e^{-\beta E_i/2}\ket{i}_1\ket{i}_2.
\ee
In general, determining the coefficients $A_{i_1,\dots,i_n}$ from the path integral over an arbitrary $\Sigma_{n,g}$ is difficult. However, the CFT dual of $\Sigma_{n,0}$ in the puncture limit where $h_i\ll 1$ was investigated in \cite{Balasubramanian:2014hda}. It was found that in this case \eqref{bdyCFT} becomes \cite{Balasubramanian:2014hda}
\be
\ket{\Sigma_{n,0}}=\sum_{i_1,\dots,i_n}C_{i_1\dots i_n}e^{-\tilde{\beta}_1 E_{i_1}/2}\dots e^{-\tilde{\beta}_n E_{i_n}/2}\ket{i_1}_1\dots\ket{i_n}_n,
\ee
where $C_{i_1\dotsi_n}$ depend on the n-point function of the CFTs and the moduli parameters,
\be
\tilde{\beta}_i=\beta_i-\log r_d-2\log 3,
\ee
$\beta_i$ is the inverse temperature of the BTZ geometry in the exterior of the $i^{\text{th}}$ asymptotic region, and $r_d$ is an undetermined constant that is independent from the moduli parameters for $(3,0)$ geometry but in general depends on the internal moduli for $n>3$ (see \cite{Balasubramanian:2014hda}).

In the hot limit, the entanglement structure of $\ket{\Sigma_{n,0}}$ was investigated in \cite{Marolf:2015vma}. In particular, it was found that the bipartite entanglement between any two CFTs at different boundaries, up to exponentially small corrections, is that of the thermofield-double state over a large region of AdS scale size\footnote{This is the same region denoted by $D_\phi$ in section \ref{sec2.3} where the distance $d_{ij}$ between the two horizons $H_i$ and $H_j$ is exponentially small.}. Thus, the CFT state dual to the local geometry in this particular region (extending between the $i^{\text{th}}$ and $j^{\text{th}}$ asymptotic regions through the causal shadow) is well approximated by $\ket{\Sigma_{2,0}}_{ij}=\ket{\text{TFD}}_{ij}$. This result will be important below in making hot multi-boundary wormholes traversable.

\section{Traversability in BTZ black holes}
\label{sec:TravReview}
In this section, we give a general review of the construction of traversable wormholes in BTZ black holes via double trace deformations \cite{Gao:2016bin}, including the case with rotation \cite{Caceres:2018ehr} and nontrivial dependence on the transverse coordinate (following \citep{Fu:2018oaq}).

In general, the perturbative construction of traversable wormholes is associated with violations of the averaged null energy condition (ANEC) along generators of Killing horizon in some classical background spacetime. We review the relation between such a violation and its perturbative backreaction on the BTZ metric below. We will then review how a double trace deformation can cause such a violation.
\subsection{Metric perturbation}
The metric of a rotating BTZ black hole in the co-rotating coordinates is obtained by substituting for the co-rotating transverse coordinate $x=\phi-\frac{r_-}{r_+}t$ into \eqref{BTZmetric} to find\footnote{In sections \ref{sec:TravReview} and \ref{sec:MultiTrav}, for simplicity, of notation we use coordinates without subscripts for the BTZ coordinates. Such coordinates should not be confused with the global AdS$_3$ coordinates of section \ref{sec:AdS3}.}
\begin{equation}
\label{corometric}
d s^{2}=-\frac{\left(r^{2}-r_{+}^{2}\right)\left(r^{2}-r_{-}^{2}\right)}{ r^{2}} d t^{2}+\frac{ r^{2}}{\left(r^{2}-r_{+}^{2}\right)\left(r^{2}-r_{-}^{2}\right)} d r^{2}+r^{2}(\mathcal{N}(r) d t+d x)^{2}
\end{equation}
where
\begin{equation}
\mathcal{N}(r)=\frac{r_{-}}{ r_{+}} \frac{r^{2}-r_{+}^{2}}{r^{2}}.
\end{equation}
\hide{
The thermodynamic quantities are given by
\begin{equation}
M=\frac{r_{+}^{2}+r_{-}^{2}}{8 G_{N} }, \quad J=\frac{r_{+} r_{-}}{4 G_{N} }, \quad S=\frac{\pi r_{+}}{2 G_{N}}, \quad \kappa=\frac{r_{+}^{2}-r_{-}^{2}}{ r_{+}}, \quad \beta=\frac{2 \pi}{\kappa}
\end{equation}
}We can pass to Kruskal coordinates by defining the right- and left-moving null coordinates. In the right exterior region, they are defined as
\begin{equation}
U=e^{\kappa u}, \quad V=-e^{-\kappa v},
\end{equation}
where $\kappa=(r_{+}^{2}-r_{-}^{2})/ r_{+}$ is the surface gravity, $u,v=t\pm r_*$ are the outgoing/ingoing coordinates, and the tortoise coordinate $r_{*}$ is
\begin{equation}
r_{*}=\frac{1}{2 \kappa} \log \frac{\sqrt{r^{2}-r_{-}^{2}}-\sqrt{r_{+}^{2}-r_{-}^{2}}}{\sqrt{r^{2}-r_{-}^{2}}+\sqrt{r_{+}^{2}-r_{-}^{2}}}.
\end{equation}
This gives the metric
\begin{equation}
\label{Kruskal}
\mathrm{d} s^{2}=\frac{1}{(1+U V)^{2}}\left\{-4  \mathrm{d} U \mathrm{d} V+4 r_{-}(U \mathrm{d} V-V \mathrm{d} U) \mathrm{d} x+\left[r_{+}^{2}(1-U V)^{2}+4 U V r_{-}^{2}\right] \mathrm{d} x^{2}\right\}.
\end{equation}
Note that the asymptotic boundary in Kruskal coordinates is located at $UV=-1$.

To linear order, the geodesic equation implies that a null ray starting from the left boundary in the far past (where $V=0$ and $U=-\infty$) satisfies
\begin{equation}
\label{Vhkk}
V(U)=-\left(2 g_{U V}(V=0)\right)^{-1} \int_{-\infty}^{U} \mathrm{d} U h_{k k}=\frac{1}{4}\int_{-\infty}^{U} \mathrm{d} U h_{k k},
\end{equation}
where $h_{kk}$ is the norm of $k^a=(\partial/\partial U)^a$ after first-order backreaction from the quantum stress tensor. To get $h_{kk}$ from the stress tensor, we use the linearized Einstein equations:
\begin{equation}
\begin{aligned}
8 \pi G_N\left\langle T_{k k}\right\rangle=&-\frac{1}{2  r_{+}^{2}}\left[\left(r_{-}^{2}-r_{+}^{2}\right) h_{k k}+2 r_{-} \partial_{x} h_{k k}+\partial_{x}^{2} h_{k k}\right.\\
&\left.+\left(r_{-}^{2}-r_{+}^{2}\right) \partial_{U}\left(U h_{k k}\right)-2  \partial_{U} \partial_{x} h_{k x}+ \partial_{U}^{2} h_{xx}\right],
\end{aligned}
\end{equation}
where $T_{kk}=T_{ab}k^a k^b$. To find the shift $\Delta V$ at $U=+\infty$, one merely needs to integrate this equation over all $U$.  This yields
\begin{equation}
\label{Tkkhkk}
8 \pi G_N \int_{-\infty}^{+\infty}\left\langle T_{k k}\right\rangle \mathrm{d} U=-\frac{1}{2  r_{+}^{2}}\left[\left(r_{-}^{2}-r_{+}^{2}\right)+2 r_{-} \partial_{x}+ \partial_{x}^{2}\right] \int_{-\infty}^{+\infty} h_{k k} \mathrm{d} U,
\end{equation}
where asymptotic AdS boundary conditions have been used.

In \citep{Gao:2016bin, Caceres:2018ehr}, the authors consider boundary couplings that are independent of the transverse coordinate for simplicity. In that case, $h_{kk}$ is independent of $x$, and equation \eqref{Tkkhkk} can be simplified to take the form
\begin{equation}
8 \pi G_N \int\left\langle T_{k k}\right\rangle \mathrm{d} U=\frac{r_{+}^{2}-r_{-}^{2}}{2  r_{+}^{2}} \int h_{k k} \mathrm{d} U,
\end{equation}
and the shift of $V$ coordinate at $U=+\infty$ is
\begin{equation}
\Delta V(+\infty)=\frac{1}{4} \int_{-\infty}^{+\infty} \mathrm{d} U h_{k k}=\frac{4\pi G_N r_+^2}{r_+^2-r_-^2}\int\left\langle T_{k k}\right\rangle \mathrm{d} U.
\end{equation}

More generally, we could consider a boundary coupling that has nontrivial dependence on the transverse coordinate. Then we could solve \eqref{Tkkhkk} using a Green's function $H$ \cite{Fu:2018oaq}
\begin{equation}
\label{greensfn}
\left(\int \mathrm{d} U h_{k k}\right)(x)=8 \pi G_N \int \mathrm{d} x^{\prime} H\left(x-x^{\prime}\right) \int \mathrm{d} U\left\langle T_{k k}\right\rangle\left(x^{\prime}\right)
\end{equation}
with
\begin{equation}
H\left(x-x'\right)=\left\{\begin{array}{cc}
\frac{r_+ e^{-(r_{+}-r_{-})(x'-x) }}{1-e^{-2 \pi(r_{+}-r_{-}) }} + \frac{r_+ e^{(r_{-}+r_{+})(x'-x) }}{ e^{2 \pi(r_{-}+r_{+})} -1} & \quad x' \geq x \\
\frac{r_+ e^{(r_{-}+r_{+})(2 \pi-x+x') }}{ e^{2 \pi(r_{-}+r_{+}) }-1}+\frac{r_+ e^{-(r_{+}-r_{-})(2 \pi-x+x') }}{1-e^{-2 \pi(r_{+}-r_{-}) }} & \quad x' \leq x
\end{array}\right.
\end{equation}
in position space where $x, x^{\prime} \in[0,2 \pi)$. In Fourier space, $H$ takes the form
\begin{equation}
H\left(x-x^{\prime}\right)=\sum_{q} e^{i q\left(x-x^{\prime}\right)} H_{q}, \quad H_{q}=\frac{1}{2 \pi} \frac{2  r_{+}^{2}}{r_{+}^{2}-r_{-}^{2}-2 i q  r_{-}+ q^{2}}.
\end{equation}
If we are working with planar BTZ black holes, $H$ takes the following form,
\begin{equation}
\label{planarH}
H\left(x-x'\right)=\left\{\begin{array}{ll}
r_+ e^{-(r_-+r_+)(x'-x)} \quad x'\geq x\\
r_+ e^{-(r_+-r_-)(x-x')} \quad x'\leq x,
\end{array}\right.
\end{equation}
where $x$ and $x'$ can take value on the whole real axis, and in Fourier space one should just adapt the sum in the compact case to an integral.

Note that, in particular, the zero-mode Green's function diverges in the extremal limit. This means that our perturbation theory breaks down in that limit, although this still suggests that the wormhole will be open for quite a long time, as will be shown below.

In contrast, the non-zero modes of $H_q$ remains finite at extremality. So in the extremal limit, it suffices to study only the zero mode. Recalling that the BTZ temperature is given by $T_H=\frac{r_{+}^{2}-r_{-}^{2}}{2 \pi r_{+} }$, we have
\begin{equation}
\frac{\pi T_H}{r_{+}} \int h_{k k} \mathrm{d} U \mathrm{d} x=8 \pi G_N \int\left\langle T_{k k}\right\rangle \mathrm{d} U \mathrm{d} x,
\end{equation}
so that \eqref{Vhkk} gives the average shift $\Delta V(U)\equiv V(U)-V(-\infty)$ as
\begin{equation}
T_H \Delta V_{\text {average }}(U)=2 G_N r_{+} \int_{-\infty}^{U} \int_{0}^{2\pi}\left\langle T_{k k}\right\rangle \mathrm{d} U \mathrm{d} x.
\end{equation}

But in any case, we could use \eqref{Vhkk} and \eqref{greensfn} to calculate the shift $\Delta V(U)$. In particular, the shift at $U=+\infty$ is given by
\begin{equation}
\label{DeltaVinf}
\Delta V(+\infty)=\frac{1}{4} \int_{-\infty}^{\infty} \mathrm{d} U h_{kk}=2\pi G_N \int \mathrm{d} x^{\prime} H\left(x-x^{\prime}\right) \int_{-\infty}^{\infty} \mathrm{d} U\left\langle T_{k k}\right\rangle\left(x^{\prime}\right).
\end{equation}

By choosing the boundary conformal frame to be $ds^2_{\partial BTZ}=-dt^2+d\phi^2=-dt^2+\left(dx+\frac{r_-}{r_+}dt\right)^2$, we can relate the boundary time with the $V$ coordinate via
\begin{equation}\label{tbdyKruskalV}
t= -\frac{ r_{+}}{r_{+}^{2}-r_{-}^{2}} \log \left(\pm V\right).
\end{equation}
Here the sign is $+$ for the left boundary and is $-$ for the right boundary.
The shortest transit time $t_*$ from left to right boundary is realized by the geodesic that leaves the left boundary at $V=-|\Delta V|/2$ and arrives at the right boundary at $|\Delta V|/2$ so that
\begin{equation}
t_{*}=-\frac{2 r_{+}}{r_{+}^{2}-r_{-}^{2}}\log \left(\frac{|\Delta V|}{2 }\right).
\end{equation}
We can also calculate the shift of the boundary angular coordinate between one end of the null geodesic and the other. Since on the horizon of the unperturbed geometry we simply follow a particular generator where $x$ is constant, on the boundary the change in $\phi$ is
\begin{equation}
\phi_*=-\frac{2 r_{-}}{r_{+}^{2}-r_{-}^{2}}\log \left(\frac{|\Delta V|}{2 }\right).
\end{equation}

\subsection{Violation of ANEC from a double trace deformation}
In AdS/CFT, the eternal BTZ black hole is dual to the thermofield double (TFD) state
\begin{equation}
\label{TFD}
|\Psi\rangle=\frac{1}{\sqrt{Z\left(\beta, \Omega_{H}\right)}} \sum_{n} e^{-\beta\left(E_{n}-\Omega_{H} J_{n}\right) / 2}\left|E_{n}, J_{n}\right\rangle_{L}\left|E_{n}, J_{n}\right\rangle_{R}.
\end{equation}
Traversability is achieved by coupling the two boundaries using a double-trace deformation
\begin{equation}\label{deltaSsamet}
\delta S= \int dtdx \ h(t,x)\mathcal{O}_{R}\left(t, x\right) \mathcal{O}_{L}\left(-t, x\right)=-\int dt\  \delta H,
\end{equation}
where $\mathcal{O}_{L/R}$ is a scalar operator living in the left/right CFT, and we choose its scaling dimension to be $\Delta=\frac{d}{2}-\sqrt{\left(\frac{d}{2}\right)^{2}+m^{2}}$ in order to have a relevant deformation \cite{Gao:2016bin}. The boundary operator $\mathcal{O}_{L/R}$ is dual to a bulk scalar field $\Phi_{L/R}$ with mass $m$.
To make the wormhole traversable, $h(t,x)$ needs to be of some definite sign for a period of time, which we denote as $[t_0,t_f]$.

We now show how such a boundary coupling leads to a violation of the ANEC. The starting point is to evaluate the bulk two-point function along the horizon $V=0$:
\begin{equation}
G\left(U, U^{\prime}\right) \equiv\left\langle \Phi_{R}(U, x) \Phi_{R}\left(U^{\prime}, x \right)\right\rangle.
\end{equation}
In a perturbative expansion in powers of the boundary coupling, the one-loop contribution to the two-point function is \cite{Gao:2016bin}
\begin{equation}
G_h=2 \sin (\pi \Delta) \int_{t_0}^{t} d t_{1}dx_1\  h\left(t_{1},x_1\right) \mathcal{K}\left(r',t',x';-t_{1}+i \beta / 2,x_1\right) \mathcal{K}_{\text{ret}}\left(r,t,x;t_1,x_1\right)+\left(t \leftrightarrow t^{\prime}\right)
\end{equation}
where $\mathcal{K}$ is the bulk-to-boundary propagator, and $\mathcal{K}_{\text{ret}}$ is the retarded bulk-to-boundary propagator. Since the BTZ black hole is just quotiented AdS$_3$, the propagators take the same form as those in AdS$_3$ but with a sum over images. The bulk-to-boundary propagator in the right exterior region in rotating BTZ metric is \cite{Gao:2016bin, Caceres:2018ehr}
\begin{equation}
\mathcal{K}\left(z, t, x ; t_{1}, x_{1}\right)=\frac{\left(r_{+}^{2}-r_{-}^{2}\right)^{\frac{\Delta}{2}}}{2^{\Delta+1} \pi } \sum_{n=-\infty}^{\infty}\left[-\sqrt{z-1} \cosh \left(\kappa \delta t-{r_{-}} \delta x_{n}\right)+\sqrt{z} \cosh \left({r_{+}} \delta x_{n}\right)\right]^{-\Delta}
\end{equation}
where
\begin{equation}
z=\frac{r^{2}-r_{-}^{2}}{r_{+}^{2}-r_{-}^{2}}, \quad \delta t=t-t_{1},  \quad \delta x_{n}=x-x_{1}+2 \pi n.
\end{equation}
We may convert this to Kruskal coordinates in the right exterior region using the relations
\begin{equation}
t=\frac{1}{2\kappa}\log\left(-\frac{U}{V}\right),\quad z=\left(\frac{1-UV}{1+UV}\right)^2.
\end{equation}
Evaluated along $V=0$, $\mathcal{K}$ becomes
\begin{equation}
\mathcal{K}(U,0,x;U_1,x_1)=\frac{\left(r_{+}^{2}-r_{-}^{2}\right)^{\frac{\Delta}{2}}}{2^{\Delta+1} \pi } \sum_{n=-\infty}^{\infty} \left[-\frac{U}{U_1}e^{-r_- \delta x_n} +\cosh \left({r_{+}} \delta x_{n}\right)\right]^{-\Delta}.
\end{equation}
The other ingredient in $G_h$ is the retarded bulk-to-boundary propagator
\begin{equation}
\mathcal{K}_{\mathrm{ret}}\left(z, t, x ; t_{1}, x_{1}\right)
=\left|\mathcal{K}\left(z, t, x ; t_{1}, x_{1}\right)\right| \theta(\delta t) \theta\left(\sqrt{z-1} \cosh \left(\kappa \delta t-{r_{-}} \delta x\right)-\sqrt{z} \cosh \left({r_{+}} \delta x\right)\right).
\end{equation}
Now we are ready to write down $G_h(U,U')$:
\begin{equation}
\label{Gh}
\begin{aligned}
G_h(U,U')=&C_0 \sum_{n=-\infty}^{\infty}\int_0^{2\pi} dx_n  \int_{U_0}^{U} \frac{dU_1}{\kappa U_1} h\left(\frac{\log(U_1)}{\kappa},x_n\right)\\
&\left[ \left( e^{-r_{-} \delta x_n} U_{1} U^{\prime}+\cosh \left(r_{+} \delta x_n\right) \right) \left(e^{-r_{-} \delta x_n} \frac{U}{U_1}- \cosh \left(r_{+} \delta x_n\right)\right)\right]^{-\Delta} \tilde{\theta}+\left(U \leftrightarrow U^{\prime}\right)
\end{aligned}
\end{equation}
where $C_0=\frac{r_+^{\Delta}\kappa^{\Delta}\sin(\pi\Delta)}{2(2^{\Delta} \pi)^2 }$,  $\tilde{\theta}=\theta \left( e^{-r_-\delta x} U-U_{1} \cosh \left(r_{+} \delta x\right) \right)$, and we have used the fact that on the right boundary $t=\frac{\log(U)}{\kappa}$. 

For planar BTZ black holes we would discard the image sum and extend the range of the $x_1$ integral to the whole real axis \cite{Gao:2016bin}. But one should not forget the constraint imposed by the $\theta$-function in the retarded propagator, which requires
\begin{equation}
e^{-r_-\delta x} U-U_{1} \cosh \left(r_{+} \delta x\right) \geq0.
\end{equation}
With the Green's function at hand, the bulk stress tensor associated with the scalar field is
\begin{equation}
\left\langle T_{\mu \nu}\right\rangle=\lim _{\mathbf{x} \rightarrow \mathbf{x}^{\prime}}\left(\partial_{\mu} \partial_{\nu} G\left(\mathbf{x}, \mathbf{x}^{\prime}\right)-\frac{1}{2} g_{\mu \nu} g^{\rho \sigma} \partial_{\rho} \partial_{\sigma} G\left(\mathbf{x}, \mathbf{x}^{\prime}\right)-\frac{1}{2} g_{\mu \nu} m^{2} G\left(\mathbf{x}, \mathbf{x}^{\prime}\right)\right).
\end{equation}
When evaluated along the horizon at $V=0$, the $g_{UU}$ component of the unperturbed metric vanishes, so to leading order we have
\begin{equation}
\label{TkkGh}
\langle T_{kk}\rangle=\lim _{U^{\prime} \rightarrow U} \partial_{U} \partial_{U^{\prime}} G_{h}\left(U, U^{\prime}\right).
\end{equation}

Finally one can compute the opening of the traversable wormhole by inserting \eqref{Gh} and \eqref{TkkGh} into \eqref{DeltaVinf}. As shown in \cite{Gao:2016bin}, the result is generally non-zero.  So for the right sign of the coupling function $h$ it will give a time-advance that makes the wormhole traversable.

\section{Traversability of multi-boundary wormholes in AdS$_3$}
\label{sec:MultiTrav}
As shown in \citep{Marolf:2015vma}, for non-rotating multi-boundary wormholes in the hot limit, the boundary state locally resembles the thermofield double state in region $D_\phi$ discussed in section \ref{sec2.3}. This could be easily generalized to rotating wormholes by adding an angular potential.  In regions that we call $D_x$ (since $x$ is a more well-defined coordinate on the horizon in the rotating case), the horizons are exponentially close to each other, and corresponding local state is exponentially close to a piece of the TFD
\begin{equation}
|\Psi\rangle=\frac{1}{\sqrt{Z\left(\beta_{TFD}, \Omega_{TFD}\right)}} \sum_{n} e^{-\beta_{TFD}\left(E_{n}-\Omega_{TFD} J_{n}\right) / 2}\left|E_{n}, J_{n}\right\rangle_{L}\left|E_{n}, J_{n}\right\rangle_{R}.
\end{equation}
Since our state is only locally TFD, the parameters $\beta_{TFD}$ and $\Omega_{TFD}$ can take any value depending on the conformal frame.  They thus should not be confused with the actual black hole inverse temperature and angular velocity. In the hot limit, one expects that such wormholes can be made traversable by the approach described in section \ref{sec:TravReview}.  We will show this below focussing on the three-boundary wormhole, and in particular on the process of traversing from boundary 1 to boundary 2.

We will first set the stage by describing and justifying the planar BTZ coordinates to be used below. In these coordinates, our calculations will be very similar to those of \cite{Gao:2016bin}. We will then show that, in the hot limit, the image sum in the Green's function is well approximated by the leading term.  This greatly simplifies our calculation. Finally, we calculate the wormhole opening with a double-trace deformation, which we require to be larger than the local thickness of the causal shadow.


\subsection{Planar BTZ coordinates and the boundary coupling}
Any BTZ black hole is locally isometric to AdS$_3$, and thus also to planar BTZ.  As a result, in any contractible region $D_x$, we may use planar BTZ coordinates to describe the spacetime. Here, we use the following planar coordinates to describe both sides of the wormhole:
\begin{equation}
ds^2=-(\tilde{r}^2-\tilde{r}_+^2)d\tilde{t}^2+\frac{d\tilde{r}^2}{\tilde{r}^2-\tilde{r}_+^2}+\tilde{r}^2 d\tilde{x}^2.
\end{equation}
We think of $\tilde{x}$ as ranging over the entire  real axis, though we are most interest in some domain that corresponds to $D_x$.  The choice of $\tilde{r}_+$ is arbitrary. The corresponding Kruskal metric is
\begin{equation}
\label{planarKruskal}
\mathrm{d} s^{2}=\frac{1}{(1+\tilde{U} \tilde{V})^{2}}\left(-4 \mathrm{d} \tilde{U} \mathrm{d} \tilde{V}+\tilde{r}_{+}^{2}(1-\tilde{U} \tilde{V})^{2} \mathrm{d} \tilde{x}^{2}\right).
\end{equation}

Although there is a causal shadow between the two horizons in the hot limit, it is exponentially small in $\ell$ and $\tilde{\ell}$ over large stretches of the horizons. So if  we put the origin of the Kruskal coordinates at the bifurcation surface of horizon 1 or 2 (or any place between them) in the region where this separation is small,
we make only an exponentially small error if we then identify the above coordinates with natural BTZ coordinates in either exterior. 
This justifies using the metric \eqref{planarKruskal} for $D_x$. We will come back to this in section \ref{sec4.3}.

Note that, in the planar BTZ metric, the horizon size parameters can be scaled arbitrarily so long  as long as we change the definition of coordinates accordingly. To be more concrete, there are two kinds of coordinate transformations that we can make (they are expressed in the ordinary angular coordinate $\phi$ for now and we will come back to the co-rotating $x$ later):
\begin{enumerate}
\item ``Adjusting the temperature" (rescaling $r_+$ and $r_-$ by the same amount):
\begin{equation}
\label{rescale}
\tilde{r}=\lambda r,\quad \tilde{t}=\frac{t}{\lambda},\quad \tilde{\phi}=\frac{\phi}{\lambda}.
\end{equation}
with the new horizon parameters $\tilde{r}_\pm=\lambda r_\pm$;
\item ``Changing the angular velocity" (changing the relative size of $r_+$ and $r_-$):
\begin{equation}
\label{boost}
\begin{aligned}
(\tilde{t},\tilde{\phi})&=(t\cosh\gamma+\phi\sinh\gamma, t\sinh\gamma+\phi\cosh\gamma)\\
\tilde{r}^{ 2}&=r^2+\tilde{r}_+^{2}-r_-^2.
\end{aligned}
\end{equation}
with the new horizon parameters $\tilde{r}_+=r_+\cosh\gamma+r_-\sinh\gamma$ and $\tilde{r}_-= r_+\sinh\gamma+r_-\cosh\gamma$. As a special case, we could set $\tilde{r}_-=0$ by choosing $\gamma=-\tanh^{-1} \frac{r_-}{r_+}$. In this case we have
\begin{equation}
\label{boost1}
\begin{aligned}
(\tilde{t},\tilde{\phi})&=\left(r_+ t-r_-\phi, r_+\phi-r_-t\right)/\sqrt{r_+^2-r_-^2}\\
\tilde{r}^{2}&=r^2-r_-^2.
\end{aligned}
\end{equation}
with $\tilde{r}_+^{2}=r_+^2-r_-^2$.
\end{enumerate}

Note that we are not changing the actual temperature and angular momentum associated with any particular global BTZ horizon (which are uniquely determined by the bulk geometry).  The point is that the above description is valid only in a contractible domain where the full global structure is not apparent.  In that domain we have described the system to good approximation as a planar BTZ black hole, for which the temperature and angular velocity depend on the choice of the boundary conformal frame and are not fixed by the bulk metric.

For simplicity, we would like to choose $\tilde{r}_-=0$ and $\tilde{r}_+$ be some fixed $\mathcal{O}(1)$ number when the $r_{+,i}$'s become large.  To clarify our notation, from here on, we use tildes to mark quantities associated with the bulk planar BTZ coordinates (for which  $\tilde{r}_-=0$), and we use symbols without tildes to refer to quantities associated with the BTZ conformal frame in some asymptotic region -- perhaps with additional labels to denote the asymptotic region of interest.

Combining \eqref{rescale} and \eqref{boost1}, the coordinate transformations we will use on boundaries 1 and 2 are
\begin{equation}
\label{rescale+boost}
\begin{aligned}
(\tilde{t},\tilde{\phi})&=\left(r_{+,i} t_i -r_{-,i} \phi_i, r_{+,i}\phi_i -r_{-,i} t_i\right)/\tilde{r}_+\\
\frac{\tilde{r}^{2}}{\tilde{r}_{+,i}^2}&=\frac{r_i^2-r_{-,i}^2}{r_{+,i}^2-r_{-,i}^2},
\end{aligned}
\end{equation}
where $i=1,2$ indicate different asymptotic regions. The above should be understood as two different coordinate transformations, one for each value of $i$. 
As a result, the two boundaries will naturally define distinct notions of `time advance' $\Delta V_1 \neq \Delta V_2$ (and also for similar quantities).

It will sometimes also be useful to consider the inverse transformation:
\begin{equation}
(t_i,\phi_i)=\frac{\tilde{r}_+}{r_{+,i}^2-r_{-,i}^2}(r_{+,i} \tilde{t}+r_{-,i} \tilde{\phi},  r_{-,i}\tilde{t}+r_{+,i}\tilde{\phi}).
\end{equation}
In terms of the co-rotating coordinates, the transformations and inverse transformations for $(t,x)$ and $(\tilde{t},\tilde{x})$ are
\begin{equation}
\label{tildecorotating1}
\tilde{t}=\frac{\kappa_i t_i-r_-x_i}{\tilde{r}_+},\quad \tilde{x}=\frac{r_{+,i}}{\tilde{r}_+}x_i=\tilde{\phi}
\end{equation}
\begin{equation}
\label{tildecorotating}
t_i=\frac{\tilde{r}_+}{\kappa_i} \left(\tilde{t}+\frac{r_-}{r_+}\tilde{x}\right),\quad x_i=\frac{\tilde{r}_+}{r_{+_i}}\tilde{x}.
\end{equation}
In particular, it will be convenient to take points on the horizons with $x=0, x_i=0$ to lie deep inside the domain $D_x$ where the separation between horizons is exponentially small.
The associated Kruskal null coordinates are related by
\begin{equation}
\label{tildekruskal}
\tilde{U}=e^{-r_{-,i} x_i}U_i,\quad \tilde{V}=e^{r_{-,i} x_i}V_i,
\end{equation}
so that at $\tilde x=0$ (where $x_i=0$) we have $\tilde U = U_i, \tilde V = V_i$.  One may interpret this as saying that we have chosen all three sets of coordinates to be associated with the same reference frame at $\tilde x=0$.

From the planar coordinates we use, it is tempting to conclude that our setup can be directly reduced to that of \cite{Gao:2016bin}, reviewed in section \ref{sec:TravReview}.  But, here, the subtlety is that the boundary coupling is not naturally defined in the conformal frame related to our bulk metric. To perform calculations, we need to first look at the conformal transformations and how they act on boundary operators.
To this end, we recall that
the boundary metric in the $i^{\text{th}}$ asymptotic region is
\begin{equation}
\label{confbdy}
ds_{i}^2=-dt_i^2+d\phi_i^2=\frac{\tilde{r}_+^2}{r_{+,i}^2-r_{-,i}^2} \left(-d\tilde{t}^2+d\tilde{\phi}^2\right).
\end{equation}

A general bi-local double-trace deformation coupling boundaries $1$ and $2$ will take the form\footnote{In contrast with section \ref{sec:TravReview} (e.g. in \eqref{deltaSsamet}) we will take the boundary times to increase toward the future on all boundaries. }
\begin{equation}
\delta S=\int dt_1 dt_2 dx_1 dx_2\ f(t_1,t_2,x_1,x_2) \mathcal{O}_1 (t_1,x_1) \mathcal{O}_2(t_2,x_2).
\end{equation}
Local couplings, analogous to those used in \cite{Gao:2016bin} are obtained by taking $f$ proportional to a delta-function.  But as opposed to the TFD case studied in \cite{Gao:2016bin}, there is no preferred natural way to identify points on boundary $1$ with points on boundary $2$. We must therefore choose some diffeomorphism $\eta$ from boundary $1$ to boundary $2$ and write
\begin{equation}
f(t_1,t_2,x_1,x_2)=h(t_1,x_1)\ \delta^{(2)} (\mathbf{x_2}-\eta (\mathbf{x_1})),
\end{equation}
where $\mathbf{x_i}=(t_i,x_i)$, $i=1,2$.   Integrating out the delta function then expresses  the coupling in terms of a single set of boundary coordinates. For computational convenience, we will choose the the functions $h$ and $\eta$ such that the double-trace deformation takes a simple form when expressed in the conformal frame associated with the tilded bulk coordinates.  In particular, we take
\begin{equation}
\label{eq:tilded}
\delta S=\int d\tilde{t} d\tilde{x}\  \tilde{h}(\tilde{t}, \tilde{x})\left(\frac{r_{+,1}^{2}-r_{-,1}^{2}}{\tilde{r}_{+}^{2}}\right)^{\frac{\Delta-1}{2}}\left(\frac{r_{+,2}^{2}-r_{-,2}^{2}}{\tilde{r}_{+}^{2}}\right)^{\frac{\Delta-1}{2}} \tilde{\mathcal{O}}_{1}(\tilde{t}, \tilde{x}) \tilde{\mathcal{O}}_{2}(\tilde{t}, \tilde{x})
\end{equation}
where $\tilde{\mathcal{O}}_{1/2}$ is the quantity $\mathcal{O}_{1/2}$ conformally transformed to the above frame. Note that the expression \eqref{eq:tilded} includes conformal factors from \eqref{confbdy} to account for the transformations of boundary operators with conformal dimension $\Delta$ as well as for the Jacobian associated with the change of integration variables.

We can also choose a simple explicit form of $\tilde{h}(\tilde{t},\tilde{x})$ that turns on at some time $\tilde{t}_0$ and turns off at some later time $\tilde{t}_f$. For example, for every $\tilde{t}$ in between we could either choose a constant (and in particular $\tilde x$-independent) coupling,
\begin{equation}
\tilde{h}(\tilde{t},\tilde{x})=h\lambda^{2-2\Delta}
\label{constantcoupling}
\end{equation}
or a Gaussian in $\tilde x$ to make it localize near some angular position $\tilde{x}_0$; i.e., for $\tilde{t}_i < \tilde{t} < \tilde{t}_f$, we may take
\begin{equation}
\tilde{h}(\tilde{t},\tilde{x})=h\lambda^{2-2\Delta}\exp\left( -\frac{\tilde{r}_+^2(\tilde{x}_1-\tilde{x}_0)^2}{\sigma^2}\right),
\label{Gaussiancoupling}
\end{equation}
where $\lambda$ is some fixed quantity with dimension of temperature and $h$ is a small and dimensionless parameter. Note that \citep{Gao:2016bin, Caceres:2018ehr} both set $\lambda$ equal to the temperature of their BTZ background.  But there is no unique temperature associated with a general multi-boundary black hole, as the temperatures of the three horizons can differ.   This is not a problem.  We are free to choose $\lambda$ in any way we like, including to choose it independent of the background, so long as long as it has the correct dimensions.

\subsection{Image sum in the hot limit}
\label{sec:imagesum}
We now show that the image sum in $G_h$ can be well approximated by keeping only the leading term. Since $G_h$ is built from two bulk-to-boundary propagators, it will be useful to study them first.

The extrapolate dictionary tells us that the bulk-to-boundary propagator in the global AdS$_3$ conformal frame can be obtained from the bulk two-point function via
\begin{equation}
\label{extrapolating}
\mathcal{K}(p,q_\partial)=\lim_{r'\rightarrow\infty}r'^\Delta G(p,q)=\lim_{r'\rightarrow\infty}r'^\Delta G(r,t,x;r',t',x').
\end{equation}
Here $p$ and $q$ are two points in the AdS$_3$ bulk.  The coordinates of $q$ are those marked with primes, and the unprimed coordinates are those of $p$.

In AdS$_3$, the two-point function for a free scalar field is given by
\begin{equation}
\label{TwopointZ}
G\left(p, q\right)=G_{\mathrm{AdS}_{3}}(Z)=\frac{1}{4 \pi}\left(Z^{2}-1\right)^{-1 / 2}\left(Z+\left(Z^{2}-1\right)^{1 / 2}\right)^{1-\Delta},
\end{equation}
where $Z=1+\frac{\sigma(p,q)}{2}$ and $\sigma(p,q)$ is the (squared) distance between $p$ and $q$ in the four dimensional embedding space (sometimes called ``chordal distance" \cite{Louko:2000tp}), and with all fractional powers of positive real numbers defined by using the positive real branch. The chordal distance is related to the geodesic distance $d(p,q)$ in AdS space by
\begin{equation}\label{sigmaEq}
\sigma(p,q)=4 \sinh ^{2}\left(\frac{d\left(p,q \right)}{2}\right).
\end{equation}

When $Z$ is large, the two-point function has the expansion
\begin{equation}
G_{A d S_{3}}\left(p,q \right)=\frac{Z^{-\Delta}}{4 \pi}\left(2^{1-\Delta}+\frac{1+\Delta}{2^{1+\Delta}} Z^{-2}+\mathcal{O}\left(Z^{-3}\right)\right).
\end{equation}
In AdS$_3$, the (unrenormalized) distance between a bulk point $p$ and a boundary point $q_\partial$ has the divergent part $\log r'$, so $G_{A d S_{3}}\left(x, x^{\prime}\right)$ decays as $(r')^{-\Delta}$. But this decay is precisely cancelled by the $(r')^{\Delta}$ in the extrapolate dictionary \eqref{extrapolating}.  As a result, the bulk-to-boundary propagator can also be obtained from the bulk-to-bulk propagator by inserting into \eqref{TwopointZ} an appropriately-renormalized (and thus finite) distance between $p$ and $q_\partial$.  According to the analysis of section \ref{sec2.2}, in the conformal frame associated with the global coordinates, this renormalized distance is defined by subtracting $\log r'$ from the unrenormalized distance.

In a general conformal frame the extrapolate dictionary becomes
\begin{equation}
\mathcal{K}=\lim_{\bar{r}'\rightarrow\infty}\bar{r}'^\Delta G(r,t,x;r',t',x')
\end{equation}
where $\bar{r}'=r' |\Omega|$ and $\Omega^2$ is the conformal factor such that the boundary metric $ds^2_{\Omega}$ satisfies $ds^2=-dt^2+d\phi^2=\Omega^2 ds^2_\Omega$. Equivalently, we could obtain the correct bulk-to-boundary propagator by inserting into \eqref{TwopointZ} an appropriately renormalized bulk-to-boundary distance associated with our conformal frame.

Since the three-boundary wormholes of section \ref{sec:AdS3} are quotients of AdS$_3$, their bulk-to-boundary propagators are given by sums of AdS$_3$ propagators over image points.  In particular, for points $p$ and $q_\partial$, we need to include AdS$_3$ propagators for the point pairs $(p, g_L q_\partial g_R^t)$, where $g_L$ and $g_R$ are any ``words" formed from the left and right generators of the quotient group $\Gamma$ used to construct the wormhole.

We would like to locate the image points  $g_L q_\partial g_R^t$ and find how they contribute to the bulk-to-boundary propagator in the hot limit. Recall from section \ref{sec2.2} that there are two types of image points: 1) points inside the same boundary diamond as $q_\partial$ and 2) points in other diamonds (i.e. outside the boundary diamond that $q_\partial$ is in). As shown in section \ref{sec2.3}, when $q_\partial$ is taken to lie at a fixed location in the largest diamond non-trivial image points in the same diamond must be exponentially close to one of the fixed points at the left or right corners of the diamond. For those in other diamonds it suffices to note that such non-trivial image diamonds are exponentially small in the hot limit.

Since all (AdS-)Cauchy slices of the wormhole spacetime lift to surfaces that run through the left and right corners of each boundary diamond, and since any bulk point $p$ can be taken to lie on a spacelike (AdS-)Cauchy surface, $p$ will have spacelike separation from points close enough to these corners.  This will in particular be true of the non-trivial images of $q_\partial$ in the hot limit. This means that we use \eqref{Lspacelike} rather than \eqref{Ltimelike} to calculate the geodesic length between $p$ and those image points.

In section \ref{sec:BTZdistance}, we calculated the geodesic distance between spacelike separated bulk and boundary points in the BTZ frame. Applying that result to our image points, we found in section \ref{sec2.3} that the geodesic distance is at least linearly large in $(\ell_i+\tilde{\ell}_i)$ in the hot limit. From \eqref{TwopointZ} and \eqref{sigmaEq} we then see that the contributions to the bulk-to-boundary propagator from the image points are exponentially suppressed, and thus that they can be ignored in the hot limit.

\subsection{Traversing the causal shadow}
\label{sec4.3}
\begin{figure}
\centering
\begin{tikzpicture}
\draw (-4,-3) to (-4,3);
\draw (4,3) to (4,-3);
\draw [very thick] (-4,3) to (4,3);
\draw [very thick] (-4,-3) to (4,-3);
\draw [dashed] (-4,3) to (3,-3);
\draw [dashed] (-4,-3) to (3,3);
\draw [dashed] (4,3) to (-3,-3);
\draw [dashed] (4,-3) to (-3,3);
\node at (-0.5,0) {\textbullet};
\node at (0.5,0) {\textbullet};
\node at (-1,0) {$H_1$};
\node at (1,0) {$H_2$};
\draw [thick, <->] (1.45,0.85) to (1,1.3);
\node at (1.6,1.3) {$\Delta V_{CS}$};
\end{tikzpicture}
\caption{The Penrose diagram of a black hole spacetime with causal shadow. In particular, this could represent the causal structure of a section that contains two asymptotic regions in the three-boundary wormhole geometry. In the figure, we mark the two bifurcation surfaces $H_1$ and $H_2$, and $\Delta V_{CS}$ caused by the causal shadow. In the hot limit that we consider in the text, $\Delta V_{CS}$ is exponentially small in $\ell$ and $\tilde{\ell}$ in region $D_x$.}
\label{fig:causalshadow}
\end{figure}
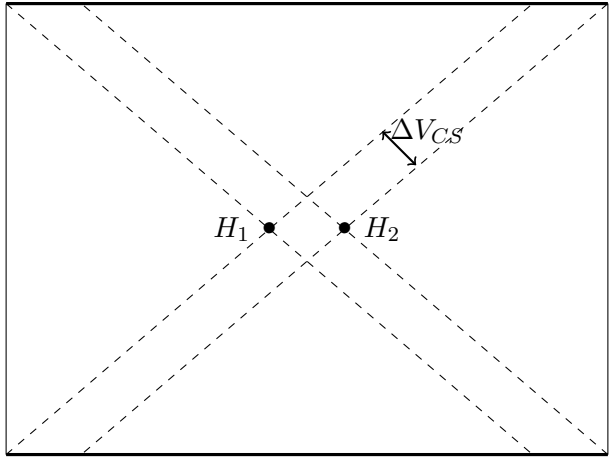

We now show in the hot limit that the $|\Delta V|$ induced by a fixed boundary coupling becomes larger than the gap $|\Delta V_{CS}|$ between horizons associated with the existence of the causal shadow region (see figure \ref{fig:causalshadow}). Thus $\Delta V_{total}\equiv |\Delta V|-|\Delta V_{CS}|$ becomes positive and therefore the wormhole is traversable.


From the above two subsections, the one-loop contribution to the Green's function is
\begin{align}\label{GhEq}
\tilde{G}_h(\tilde{U},\tilde{U'})&=\tilde{C}_0 \int d\tilde{x}_1 \int_{\tilde{U}_0}^{\tilde{U}} \frac{d\tilde{U}_1}{\tilde{r}_+ \tilde{U}_1} \tilde{h}\left(\frac{\log \tilde{U}_1}{\tilde{r}_+},\tilde{x}_1\right) \left[\left( \tilde{U}_{1} \tilde{U}'+\cosh \left(\tilde{r}_{+} \delta \tilde{x}\right)\right)\left( \frac{\tilde{U}}{\tilde{U}_{1}}-\cosh \left(\tilde{r}_{+} \delta \tilde{x}\right)\right)\right]^{-\Delta}\nonumber\\ &+ (\tilde{U} \leftrightarrow \tilde{U}'),
\end{align}
where $\tilde{U}_0=e^{\tilde{r}_+ \tilde{t}_0}$, $\delta \tilde{x}=\tilde{x}-\tilde{x}_1$
and
\begin{equation}
\tilde{C}_{0}=\frac{\tilde{r}_{+}^{2 \Delta} \sin (\pi \Delta)}{2\left(2^{\Delta} \pi\right)^{2}}\left(\frac{r_{+,1}^{2}-r_{-,1}^{2}}{\tilde{r}_{+}^{2}}\right)^{\frac{\Delta-1}{2}}\left(\frac{r_{+,2}^{2}-r_{-,2}^{2}}{\tilde{r}_{+}^{2}}\right)^{\frac{\Delta-1}{2}}=\frac{\tilde{r}_+^2(r_{+,1}^{2}-r_{-,1}^{2})^{\frac{\Delta-1}{2}}(r_{+,2}^{2}-r_{-,2}^{2})^{\frac{\Delta-1}{2}}\sin (\pi \Delta)}{2\left(2^{\Delta} \pi\right)^{2}}.
\end{equation}
The limits of the $\tilde{x}$ integral above are set by the theta function $\theta \left( \frac{\tilde{U}}{\tilde{U}_{1}}-\cosh \left(\tilde{r}_{+} \delta \tilde{x}\right)\right)$.

We can use the above result to calculate the stress tensor:
\begin{equation}
\langle \tilde{T}_{kk} \rangle =\lim_{\tilde{U}' \rightarrow\tilde{U}} \partial_{\tilde{U}'} \partial_{\tilde{U}} \tilde{G}_h (\tilde{U},\tilde{U}').
\end{equation}
If the background was exactly planar BTZ, then the shift of $\tilde{V}$ coordinate at $\tilde{U}=+\infty $ would be
\begin{equation}
\Delta \tilde{V}(\tilde{x})=2\pi G_N \int_{-\infty}^{+\infty} \mathrm{d} \tilde{x}' \tilde{H}\left(\tilde{x}-\tilde{x}' \right) \left( \int_{-\infty}^{\infty} \mathrm{d} \tilde{U} \langle \tilde{T}_{kk} \rangle\right)(\tilde{x}'),
\label{DeltaVtilde}
\end{equation}
where $\tilde{H}(\tilde{x}-\tilde{x}')$ is the Green's function \eqref{planarH} for non-compact $\tilde{x}$ and $\tilde{x}'$ when $\tilde{r}_-=0$,
\begin{equation}
H\left(\tilde{x}-\tilde{x}'\right)=
\tilde{r}_+ e^{-\tilde{r}_+|\tilde{x}'-\tilde{x}|}.
\end{equation}

From our arguments above, using this result with \eqref{rescale+boost} also gives the correct result in our three-boundary wormhole up to two sorts of corrections. The first are due to errors in \eqref{rescale+boost} associated with the finite-but-small thickness of the causal shadow, and the second comes from neglecting the sum over non-trivial images of $q_\partial$.  But both sorts of corrections are exponentially small in the hot limit as discussed above.  Thus to good approximation in the coordinates related to the $i^{\text{th}}$ boundary we find the shift $\Delta V_i$ to be
\begin{equation}
\label{untildeV}
\Delta V_i(x_i)= e^{-r_{-,i}x_i} \Delta \tilde{V}(\tilde{x}).
\end{equation}

To put this all together, recall that we are most interested in the region near $\tilde x =0$ where the separation between the bifurcation surfaces is exponentially small.  There $V_i \approx \tilde V$, and the three coordinate systems are all associated with the same frame of reference.  In particular, both bifurcation surfaces will have $U_1 + V_1 \approx constant$ and also $U_2 + V_2 \approx constant$.  Thus the exponentially small separation is also associated with exponentially small sized $\Delta \tilde V_{CS} \approx \Delta V_{1,CS} \approx \Delta V_{2,CS}$ of the causal shadow in this region.

On the other hand, near $x_i=0$ the time advance  $\Delta V_i$ is {\it not} exponentially suppressed at large $\ell_i$ and $\tilde{\ell}_i$. Instead, it has at most a polynomial suppression. Thus at large $\ell_i, \tilde \ell_i$ we find $\Delta V_i \gg \Delta V_{i, CS}$  near $\tilde x=0$ and the wormhole becomes traversable in this region.

As a consistency check, we now show that the physical quantity $\Delta V_i$ does not depend on the fictitious parameter $\tilde{r}_+$ that we have been using to simplify the calculations.
Our starting point is \eqref{GhEq}. We write $\tilde{G}_h\equiv F+F'$ where $F$ is the term explicitly shown in \eqref{GhEq}
\begin{equation}
F(\tilde{U},\tilde{U'})=\tilde{C}_0 \int d\tilde{x}_1 \int_{\tilde{U}_0}^{\tilde{U}} \frac{d\tilde{U}_1}{\tilde{r}_+ \tilde{U}_1} \tilde{h}\left(\frac{\log \tilde{U}_1}{\tilde{r}_+},\tilde{x}_1\right) \left[\left( \tilde{U}_{1} \tilde{U}'+\cosh \left(\tilde{r}_{+} \delta \tilde{x}\right)\right)\left( \frac{\tilde{U}}{\tilde{U}_{1}}-\cosh \left(\tilde{r}_{+} \delta \tilde{x}\right)\right)\right]^{-\Delta}
\end{equation}
and $F'$ is the term with $\tilde{U}$ and $\tilde{U}'$ exchanged.  Using this symmetry we may write $\langle \tilde{T}_{kk}\rangle$ in the form
\begin{equation}
\langle \tilde{T}_{kk} \rangle =2 \lim_{\tilde{U}' \rightarrow\tilde{U}} \partial_{\tilde{U}'} \partial_{\tilde{U}} F (\tilde{U},\tilde{U}').
\end{equation}

Next, we change the integration variables to make the dependence on $\tilde{r}_+$ clear. First we define a new integration variable $y\equiv\cosh (\tilde{r}_+ \delta \tilde{x})=\cosh [\tilde{r}_+ (\tilde{x}-\tilde{x}_1)]$ to write $F$ as
\begin{equation}
F(\tilde{U},\tilde{U'})=\frac{2\tilde{C}_0}{\tilde{r}_+^2}  \int_{\tilde{U}_0}^{\tilde{U}}  \frac{d\tilde{U}_1}{\tilde{U}_1} \int_1^{\tilde{U}/\tilde{U}_1} \frac{dy}{\sqrt{y^2-1}}  \tilde{h}\left(\frac{\log \tilde{U}_1}{\tilde{r}_+},\tilde{x}_1 \right)\left[\left( \tilde{U}_{1} \tilde{U}'+y \right)\left( \frac{\tilde{U}}{\tilde{U}_{1}}-y \right)\right]^{-\Delta},
\end{equation}
where the limits of the $y$ integral are determined by the theta function $\theta\left( \frac{\tilde{U}}{\tilde{U}_{1}}-\cosh \left(\tilde{r}_{+} \delta \tilde{x}\right)\right)$, and the argument $\tilde{x}_1$ in the function $\tilde{h}$ should be implicitly treated as a function of $y$.

As we can see, all the $\tilde{r}_+$ dependence in the prefactor $\frac{2\tilde{C}_0}{\tilde{r}_+^2}$ cancels out. Recall also the relations \eqref{tildecorotating1}
\begin{equation}\label{tildecorotating1New}
\tilde{r}_+ \tilde{x}=r_{+,i}x_i,\quad \tilde{r}_+ \tilde{t}= \kappa_{i} t_{i}-r_{-,i} x_{i},
\end{equation}
so that on the horizon $V=0$ we have
\begin{equation}
\tilde{U}=e^{\tilde{r}_+ \tilde{t}}=e^{\kappa_{i} t_{i}-r_{-,i} x_{i}}.
\end{equation}
Similar relations hold for $ \tilde{U},\tilde{U}'$ and $\tilde{U}_0$ in the integration limits, and they can be expressed in terms of purely boundary quantities.  Furthermore, we should avoid introducing any $\tilde{r}_+$ dependence in $\tilde{h}$ by hand. This means that, when choosing the form of $\tilde{h}$, the argument $\tilde{t}_1$ and $\tilde{x}_1$ in $\tilde{h}$ should both come with a factor of $\tilde{r}_+$, since the combination $\tilde{r}_+ \tilde{t}_1$ and $\tilde{r}_+ \tilde{x}_1$ can be converted by \eqref{tildecorotating1New} to something that only involves parameters and coordinates related to some boundary.  In terms of the new variable $y$, this means that we must have the combination $(\tilde{r}_+ \tilde{x} -\cosh^{-1} y)$ independent of $\tilde{r}_+$. Therefore, $F$ is also independent of $\tilde{r}_+$.

The physical observable $\Delta V_i$ on one boundary is
\begin{equation}
\label{appdeltav}
\Delta V_i(x_i)=e^{-r_{-,i}x_i}\ 2\pi G_N \int_{-\infty}^{+\infty} \mathrm{d} \tilde{x}^{\prime}\tilde{r}_+ e^{-\tilde{r}_+|\tilde{x}'-\tilde{x}|} \left(\int_{-\infty}^{\infty} \mathrm{d} \tilde{U}\left\langle\tilde{T}_{k k}\right\rangle\right)\left(\tilde{x}^{\prime}\right).
\end{equation}
 No dependence on $\tilde{r}_+$ is introduced in passing from $F$ to $\int \mathrm{d} \tilde{U}\langle\tilde{T}_{k k}\rangle$ and, from our previous argument, $\int \mathrm{d} \tilde{U}\langle\tilde{T}_{k k}\rangle$ as a function of $\tilde{x}'$ should only depend on the combination $\tilde{r}_+ \tilde{x}'$. As we can see, all other parts involving tilded coordinates in \eqref{appdeltav} all come with a factor of $\tilde{r}_+$, so the physical quantity $\Delta V_i$ will not have any $\tilde{r}_+$ dependence.

\subsection{Numerical results}

\begin{figure}
\begin{minipage}{0.5\textwidth}
\centering
\includegraphics[scale=0.5]{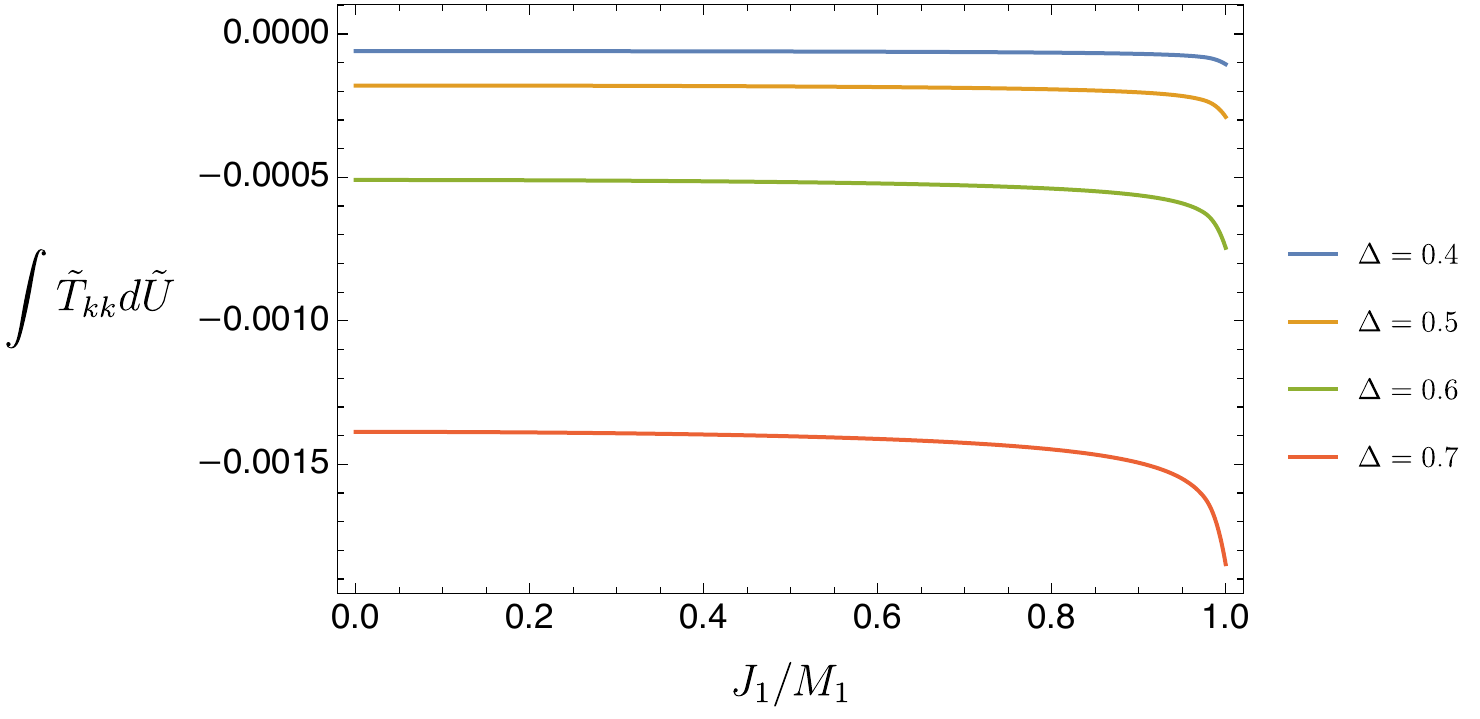}
\end{minipage}
\begin{minipage}{0.5\textwidth}
\centering
\includegraphics[scale=0.5]{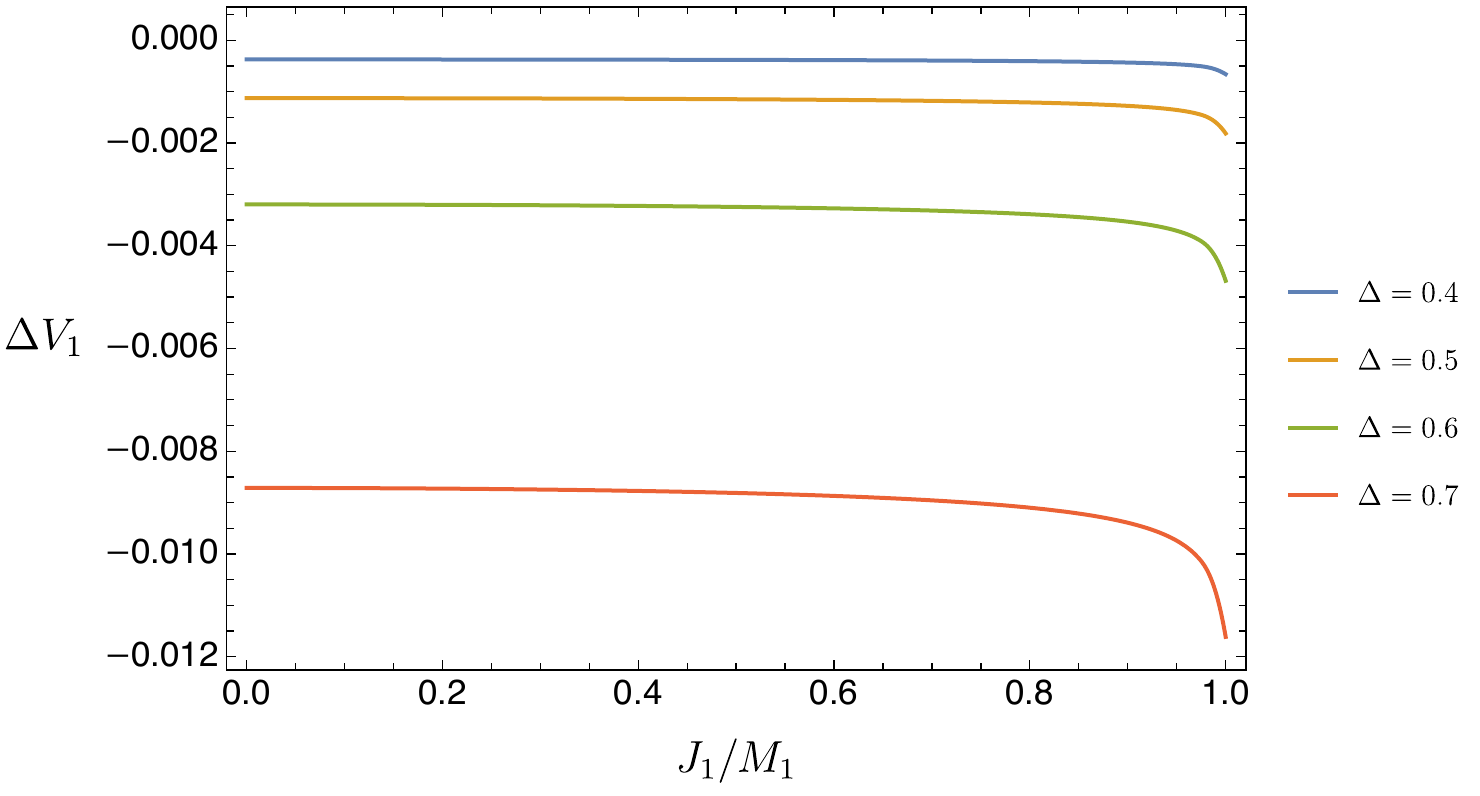}
\end{minipage}
\caption{For the case of constant coupling, the averaged null energy $\int \tilde{T}_{kk} d\tilde{U}$ (left) and the horizon shift $\Delta V_1$ at $x_1=0$ (right). In both panels, we choose $h=1$, $\lambda=1$, $G_N=1$, $r_{+,2}=100$ , $r_{-,2}=20$ and $r_{+,1}=100$.}
\label{fig:constantcoupling}
\end{figure}

\begin{figure}
\begin{minipage}{0.5\textwidth}
\centering
\includegraphics[scale=0.5]{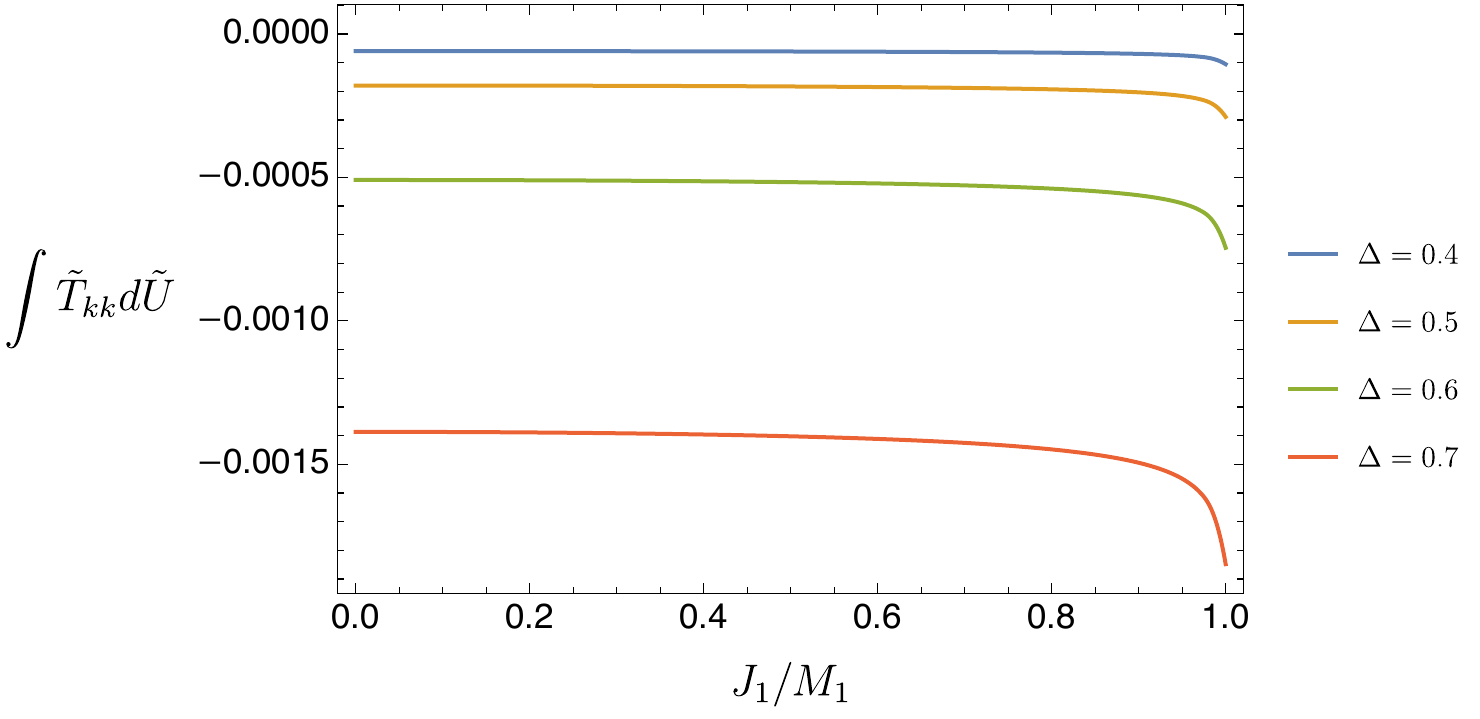}
\end{minipage}
\begin{minipage}{0.5\textwidth}
\centering
\includegraphics[scale=0.5]{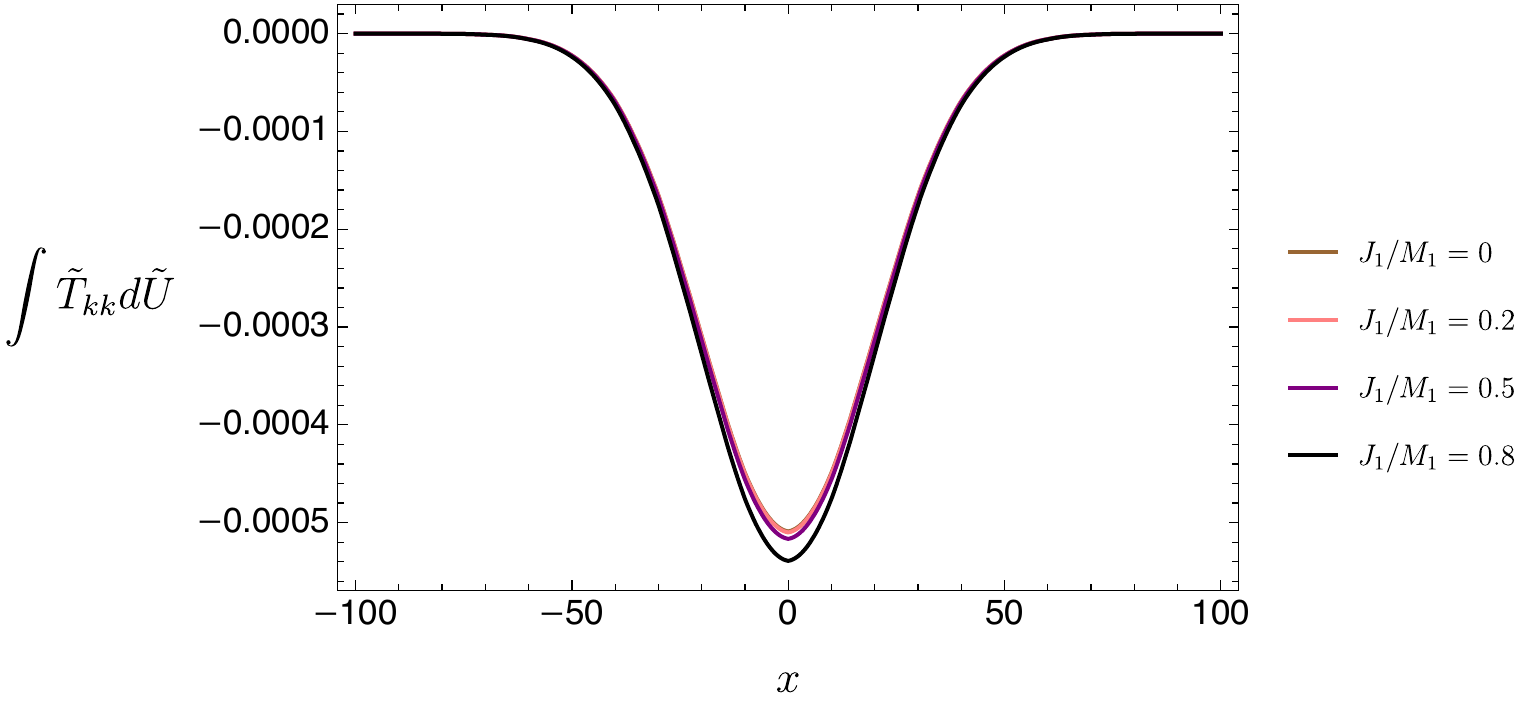}
\end{minipage}
\caption{For the case of Gaussian coupling, the averaged null energy $\int \tilde{T}_{kk} d\tilde{U}$ at $x_1=0$ (left) and its profile for general $x_1$ (right). In both panels, we choose $h=1$, $\lambda=1$, $G_N=1$ $r_{+,2}=100$ , $r_{-,2}=20$ and $r_{+,1}=100$, $\sigma=0.2$ and $x_0=0$. In the right panel we also choose $\Delta=0.6$.}
\label{fig:GaussianTUU}
\end{figure}

\begin{figure}
\begin{minipage}{0.5\textwidth}
\centering
\includegraphics[scale=0.35]{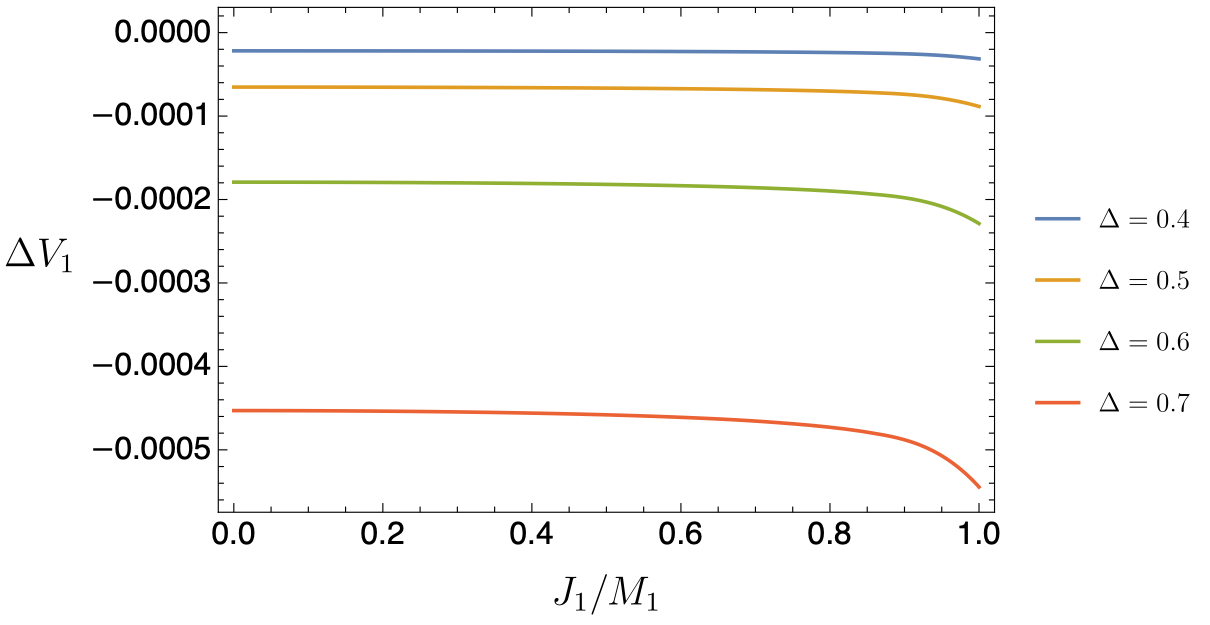}
\end{minipage}
\begin{minipage}{0.5\textwidth}
\centering
\includegraphics[scale=0.47]{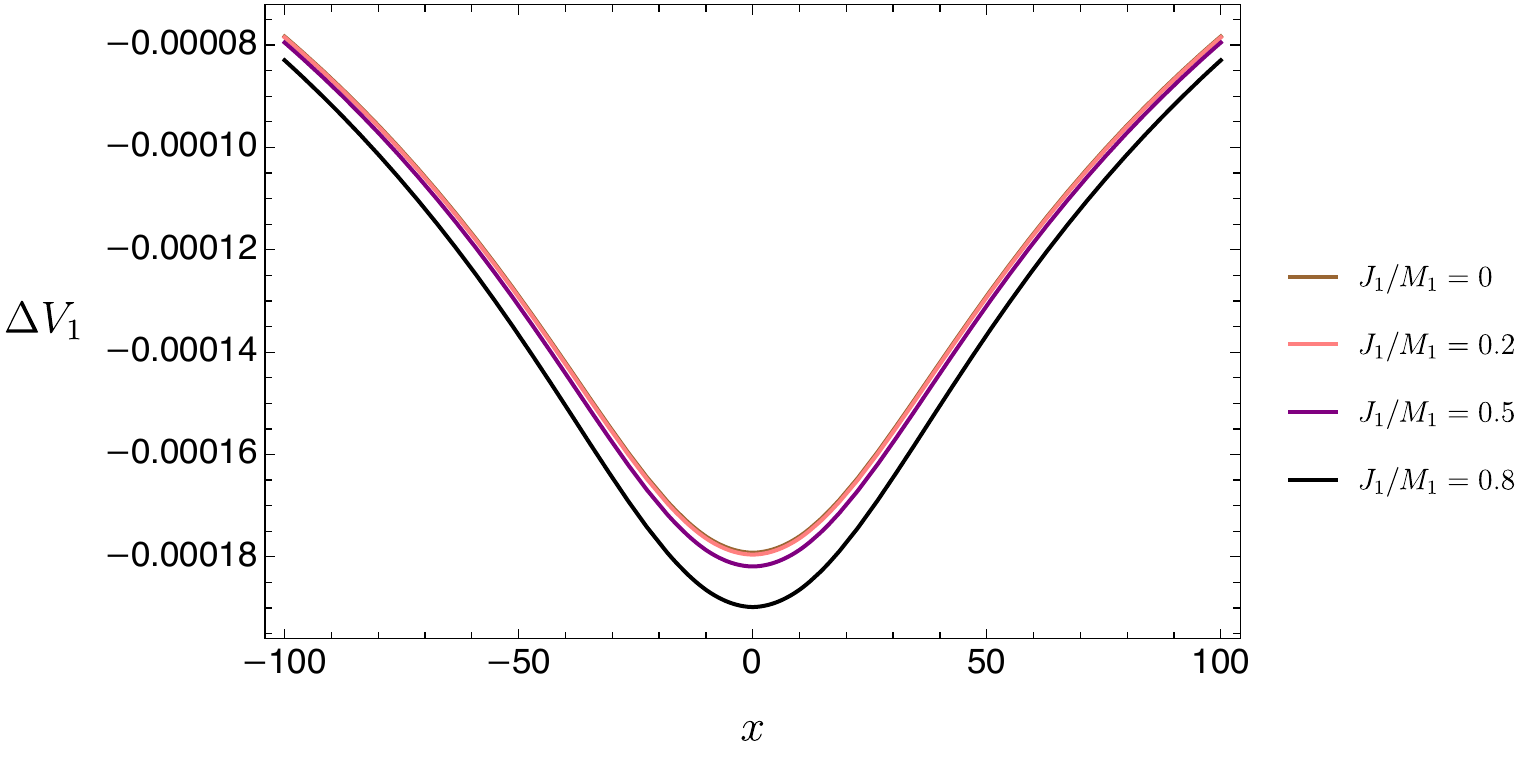}
\end{minipage}
\caption{For the case of Gaussian coupling, the shift of horizon $\Delta V_1$ at $x_1=0$ (left) and its profile for general $x_1$ (right). In both panels, we choose $h=1$, $\lambda=1$, $G_N=1$ $r_{+,2}=100$ , $r_{-,2}=20$ and $r_{+,1}=100$, $\sigma=0.2$ and $\tilde{x}_0=0$. In the right panel we also choose $\Delta=0.6$.}
\label{fig:GaussianDeltaV}
\end{figure}

We now present some numerical results in order to illustrate our construction.
Here we will take the boundary coupling to be turned on at $\tilde{t}_0=0$ and never shut off.
We will consider two types of boundary coupling: 1) for every $\tilde{t} >0$ the coupling is constant, as in \eqref{constantcoupling} and 2)  for every $\tilde{t} >0$ we take the coupling to be a Gaussian centered at some point, as in \eqref{Gaussiancoupling}.

We also take $h=1$ and $\lambda=1$ in the boundary coupling, and $G_N=1$ for simplicity. Furthermore, without loss of generality, we only consider a subspace of the wormhole parameter space defined by $r_{+,2}=100$ , $r_{-,2}=20$ and $r_{+,1}=100$.  We then study the dependence of various quantities on the remaining parameters $r_{-,1}$ (or equivalently the ratio between angular momentum and mass $J_1/M_1$ on boundary 1) and the scaling dimension $\Delta$.

The quantities studied below are the averaged null energy $\int \tilde{T}_{kk} d\tilde{U}$ and the shift of the horizon $\Delta V_1$ as measured on boundary 1. Note that here $\int \tilde{T}_{kk} d\tilde{U}$ is not a physical quantity since we could choose any kind of ``tilded coordinates", but we show it here because its negativity is important for traversability. For convenience we choose $\tilde{r}_+=1$.

Results for the case of constant coupling are shown in figure \ref{fig:constantcoupling}. There we show $\int \tilde{T}_{kk} d\tilde{U}$ and $\Delta V_1$ at $x_1=0$ (or equivalently $\tilde{x}=0$) for different $\Delta$ and $J_1/M_1$. As we can see, both quantities are negative and diverge near extremality.

For Gaussian coupling, we choose $\sigma=0.2$ and $\tilde{x}_0=0$. In figure \ref{fig:GaussianTUU} we show $\int \tilde{T}_{kk} d\tilde{U}$ at $x_1=0$ (or equivalently $\tilde{x}=0$) and its angular dependence for some choices of parameters, while results about $\Delta V_1$ are shown in figure \ref{fig:GaussianDeltaV}.

\section{Discussion}
\label{sec:discussion}
The above work extends the Gao-Jafferis-Wall traversability protocol \cite{Gao:2016bin} to multi-boundary wormholes. The main physical difficulty in achieving traversability in this case is the existence of the causal shadow region between the horizons, and the main technical complication in the analysis involves calculating the image sum in the Green's function. Our main result is that, in the hot limit, both of these difficulties can be circumvented and traverseability can be demonstrated for appropriate couplings. As shown in section \ref{sec:AdS3}, this is because for any pair of horizons there is a region whose extent along the horizons  is large in comparison with the AdS length where the horizons are exponentially close to each other.   The analysis in such regions thus reduces to that of \cite{Gao:2016bin}.  In particular, in this limit the distance between the global AdS$_3$ images of appropriate bulk points bceomes large, which exponentially suppresses all but one of the corresponding contributions to the Green's function relative to the largest such contribution.   This greatly simplifies the calculation of the Green's function required to calculate the average null energy along the horizon.  In a dual field theory description, the essential point is that the CFT
state in this region is approximately given by the TFD state \cite{Marolf:2015vma}.

\begin{figure}[t]
\centering
\begin{tikzpicture}
\draw[black,thick,dashed] (2,0) ellipse (0.15 and 0.55);
\draw[black,thick,dashed] (-2,0) ellipse (0.15 and 0.55);
\draw[black,thick,solid] (0.15,0.1) arc (20:-200:0.3);
\draw[black,thick,solid] (-0.48,-0.025) arc (-195:-345:0.36);
\draw[black,thick,solid] (-3,1) to[out=-40,in=180] (-2,0.55) to[out=0,in=180] (-0.1,1.45) to[out=0,in=180] (2,0.55)  to[out=0,in=-140] (3,1);
\draw[black,thick,solid] (-3,-1) to[out=40,in=180] (-2,-0.55) to[out=0,in=180] (-0.1,-1.45) to[out=0,in=180] (2,-0.55)  to[out=0,in=140] (3,-1);
\draw[black,thick,dotted] (-0.1,0.85) ellipse (0.1 and 0.6);
\draw[black,thick,dotted] (-0.1,-0.89) ellipse (0.1 and 0.57);
\draw[black,thick,solid] (3.0,0) ellipse (0.1 and 1.0);
\draw[black,thick,solid] (-3.0,0) ellipse (0.1 and 1.0);
\end{tikzpicture}
\caption{A Cauchy slice of the $(2,1)$ geometry showing the horizons (dashed lines) and the two extremal surfaces (dotted lines) in the causal shadow region. In the hot limit, the length of both types of surfaces have to be taken to be large so that, by the Gauss-Bonnet theorem, there will be a large region where they are arbitrarily close to each other.}
\label{fig:genusCase}
\end{figure}
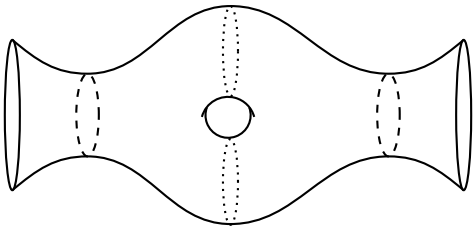

Although we presented explicit calculations only for the three-boundary wormhole geometry, our work can be generalized to general $n$-boundary genus $g$ wormholes  (i.e. to $(n,g)$ geometries).   The one subtlety in doing so is that, in addition to taking a hot limit for the horizons, one must also take similar limits of certain internal moduli in order to make the causal shadow become thin.
See figure \ref{fig:genusCase} for the case $n=2$, $g=1$, but similar issues arise even for $g=0$ when $n>3$.  Indeed, one can view this as a result of the fact that a general $(n,g)$ geometry can be made by sewing together copies of $(3,0)$ ``pair of pants" geometries, but that in doing so some of the minimal circles that would have defined horizons in some given $(3,0)$ geometry become cycles inside the causal shadow of the final $(n,g)$ geometry.  Thus, the desired hot limit involves not only taking limits of the parameters that define the final $(n,g)$ horizons, but also requires us to take limits of the parameters associated with the would-be $(3,0)$ horizons that are now inside the causal shadow.   That this is possible in general was shown in \cite{Marolf:2015vma} for the static case, but those arguments can be generalized to allow rotation just as in section \ref{sec:AdS3} above.  Thus the traversability analysis reduces to exactly the same one we used for the case without genus,  and once again the CFT dual to the bulk region where the horizons are exponentially close together is well-approximated by the TFD state.

In the extremal limit, we showed in appendix \ref{appdij} that the minimal distance $d_{ij}$ between the horizons diverges logarithmically. However, from \eqref{GhEq} and \eqref{DeltaVtilde}, we see that the time advance $\Delta V$ induced by the double trace deformation diverges polynomially, which is also illustrated in figures \ref{fig:constantcoupling} and \ref{fig:GaussianDeltaV}. For this reason, we expect that the wormhole is still traversable in the extremal limit even though, as discussed in section \ref{sec:TravReview}, the perturbative analysis that allowed us to calculate $\Delta V$ will no longer be valid\footnote{For further discussion on traversable wormholes in the extremal limit, see \cite{Fallows:2020ugr}.}.


Recall that, in the ER=EPR proposal \cite{Maldacena:2013xja}, entanglement between two (non-interacting) quantum systems is geometrically realized by a non-traversable wormhole (i.e. Einstein-Rosen bridge)  connecting them. When the two systems are allowed to dynamically interact with each other via a quantum interaction like the double trace deformation, a quantum teleportation protocol becomes possible and quantum information can be teleported between them through the wormhole that now becomes traversable. As pointed out in \cite{Gao:2016bin}, this is distinct from the standard quantum teleportation protocol where only classical interactions are allowed between the two entangled systems (though see \cite{Susskind:2017nto} for connections with standard quantum teleportation). On the one hand, this provided a concrete mechanism for recovery of quantum information via the Hayden-Preskill protocol \cite{Hayden:2007cs} from the Hawking radiation of old black holes \cite{Maldacena:2017axo}. One the other hand, it inspired a number of experimental proposals (e.g. \cite{Yoshida:2018vly,Brown:2019hmk}) for quantum teleportation via quantum interactions between two entangled systems\footnote{The proposal \cite{Yoshida:2018vly} was experimentally realized in \cite{Landsman:2018jpm} using an ion trap quantum computer.}. Looked at from this perspective, and although our construction holds in the limit where the mulitpartite entanglement is ignored, 
our work is a first step toward a generalization of the quantum teleportation protocol to quantum systems with multipartite entanglement. Since the CFT state dual to a general $(n,g)$ geometry is not known for general values of the moduli parameters, one can focus on the hot limit where locally the entanglement is mainly bipartite and is approximately a TFD state. It would be interesting to realize such a quantum state in the lab and perform the quantum teleportation protocol on it. As discussed in this work, the main new features in this case are the causal shadow region as well as the non-trivial angular dependence. It would be interesting to understand how these features are realized in an experimental set-up of quantum teleportation in the case of quantum circuits with multipartite entanglement. We expect that, in this case, the traversability protocol will occur on a mixed TFD state and that the ``size" of the causal shadow region will provide an upper bound on the fidelity of the teleported state.   See also  \cite{Emparan:2020ldj} for a 3-mouth traversable wormhole where multipartite entanglement may play a larger role.

 As discussed in \cite{Maldacena:2017axo}, the experience of an observer passing through a two-sided traversable wormhole is that of a smooth free fall through a low-curvature spacetime. For an observer entering a multi-boundary wormhole, the experience will be similarly pleasant only for particular angular domains. Entering the wormhole from other directions will require the observer to become trapped inside the black hole and to reach the singularity.  One should thus be sure of the accuracy of one's trajectory when entering such a wormhole.

There are several directions for future investigations. First, it would be interesting to extend this work to higher dimensions, where gravity is more interesting than in three dimensions.
In addition, as discussed above, this work can be interpreted as a quantum teleportation circuit with multipartite entanglement as a resource. Therefore, one can extend the analysis of \cite{Yoshida:2018vly,Brown:2019hmk} to this case and characterize how multipartite entanglement affects the properties and conditions of teleportation.

\paragraph{Acknowledgments}
AA and DM would like to thank the Kavli Institute for Theoretical Physics for its hospitality during a portion of this work. As a result, this research was supported in part by the National Science Foundation under Grant No. NSF PHY-1748958. AA acknowledges funding from the Institute of Particle Physics, Canada, through an Early Career Theory Fellowship.  ZW and DM were supported by NSF grant PHY-1801805 and funds from the University of California.

\appendix
\section{An alternative construction of the three-boundary black hole}\label{appSymm}
We constructed a three-boundary black hole in section \ref{sec2.1} by choosing some AdS$_3$ isometries and taking a quotient by the group $\Gamma$ that they generate. Although the representation of the generators used there is convenient for calculation, it makes the third asymptotic region (whose horizon is generated by $\gamma_1^{-1}\gamma_2$ and $\gamma_1\gamma_2^{-1}$) appear to be on a different footing than the other two. In particular, as described in the standard AdS$_3$ conformal frame the coordinate size of this third region vanishes in the hot limit. To show that this is an artifact of our choice of generators, we give an alternative representation below where the coordinate size of the third boundary is non-vanishing in the hot limit. For simplicity, we focus on the non-rotating case which is generated by a diagonal subgroup of isometries where $\gamma_L=\gamma_R\equiv \gamma$.  Dropping this diagonal restriction will give a generalized to the rotating case.

We begin with the most general form of a $SL(2,\mathbb{R})$ generator:
\begin{equation}
\xi=x_1J_1+x_2J_2+x_3J_3.
\end{equation}
This generator is hyperbolic when $x_1^2+x_2^2-x_3^2>0$, which is equivalent to the requirement $\Tr e^\xi>2$. The length of horizon generated by $\gamma=e^\xi$ is
\begin{equation}
\ell=2\cosh^{-1} \frac{\Tr \gamma}{2}=\sqrt{x_1^2+x_2^2-x_3^2}.
\end{equation}
It is thus natural to parametrize our generator as
\begin{equation}
\xi=\ell (\cosh\alpha\sin\beta J_1 +\cosh\alpha\cos\beta J_2 - \sinh\alpha J_3)\equiv \ell (\vec{a}\cdot\vec{J})
\end{equation}
where the generator is written as an inner product  taken with signature $(++-)$, where $\vec{a}=(\cosh\alpha\sin\beta,\cosh\alpha\cos\beta,\sinh\alpha)$, and $\vec{J}=(J_1,J_2,J_3)$.

To make a three-boundary wormhole, we choose two such generators
\begin{equation}
\xi_1=\ell_1 (\cosh\alpha_1\sin\beta_1 J_1 +\cosh\alpha_1\cos\beta_1  J_2 - \sinh\alpha_1  J_3)=\ell_1 (\vec{a}_1\cdot\vec{J})
\end{equation}
\begin{equation}
\xi_2=\ell_2 (\cosh\alpha_2\sin\beta_2 J_1 +\cosh\alpha_2\cos\beta_2  J_2 - \sinh\alpha_2  J_3)=\ell_2 (\vec{a}_2 \cdot\vec{J}),
\end{equation}
so that the corresponding group elements are $\gamma_1=e^{\xi_1}$ and $\gamma_2=e^{\xi_2}$. Then the group element related to the third asymptotic region is $\gamma_3=- \gamma_1^{-1}\gamma_2$. As a result, the horizon length of the third region are related to our parameters by
\begin{equation}
\cosh \frac{\ell_3}{2}=-\cosh \frac{\ell_1}{2}\cosh \frac{\ell_2}{2}+\sinh \frac{\ell_1}{2}\sinh \frac{\ell_2}{2} (\vec{a}_1\cdot \vec{a}_2).
\end{equation}
Note that our geometry depends only on the three parameters $\{\ell_1, \ell_2, \vec{a}_1\cdot \vec{a}_2\}$, or equivalently $\{\ell_1, \ell_2, \ell_3\}$.  This gives the expected three-dimensional moduli space for a non-rotating 3-boundary wormhole.

Our previous representation corresponds to the choice $\vec{a}_1=(0,-1,0)$ and $\vec{a}_2=(0,-\cosh\alpha,-\sinh\alpha)$. These choices reproduce our previous results.  In particular, our previous representation does not involve $J_1$.

\begin{figure}
\centering
\begin{tikzpicture}
\draw (0,0) circle[radius=3];
\draw [dashed] (-0.025,3) to [out=270,in=0] (-3,0);
\draw [dashed] (0.025,3) to [out=270,in=180] (3,0);
\draw [dashed] (-3,-0.05) to (3,-0.05);
\node[text width=0.8cm] at (1.7,1.0) {$H_1$};
\node[text width=0.8cm] at (-1.5,1.0) {$H_2$};
\node[text width=0.8cm] at (0.2,-0.4) {$H_3$};
\end{tikzpicture}
\caption{The three-boundary wormhole in the hot limit under our alternative construction with $\alpha_1=-\alpha_2=\alpha, \beta_1=-\beta_2=\beta=\frac{\pi}{4}$, where $H_1$, $H_2$ and $H_3$ are the three horizons. The fixed points of distinct generators become close to each other in this limit, but each asymptotic region remains a finite size.}
\label{fig:altrep}
\end{figure}
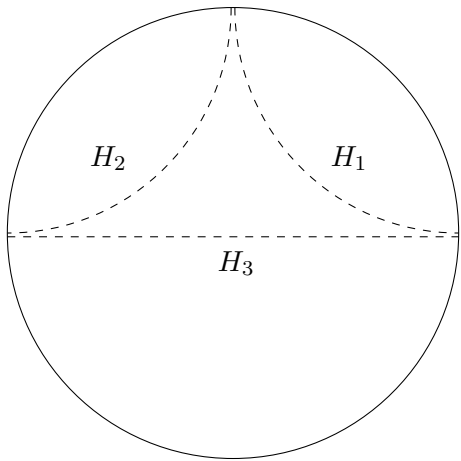

However, this choice is far from unique.  The only real restriction on the form of the generators is that the geometry not  become the one-boundary torus wormhole described in \cite{Aminneborg:1998si}. To make a $(3,0)$ wormhole, the bulk geodesic connecting the fixed points of $\gamma_1$ must not cross that connecting the fixed points of $\gamma_2$, while they cross each other in the $(1,1)$ wormhole construction.

To be definite, let us choose generators with
\begin{equation}
\alpha_1=-\alpha_2=\alpha,\quad \beta_1=-\beta_2=\beta=\frac{\pi}{4}.
\end{equation}
This ansatz still allows the freedom to vary the horizon lengths by tuning ${\ell_1,\ell_2,\alpha}$. Then, as we did in section \ref{sec2.2}, we could calculate the eigenvectors of the $\gamma_i$'s and analyze the fixed points on the boundary in the hot limit, and those fixed points are also endpoints of the horizons. For the non-rotating case, all the fixed points are on the $t=0$ slice, and here we take $\phi \in [0,2\pi)$. As shown in figure \ref{fig:altrep}, in the hot limit the endpoints of $H_1$ approach $\phi=3\pi/2$ and $\phi=\pi$, while the endpoints of $H_2$ approach $\phi=3\pi/2$ and $\phi=0$, and ethe ndpoints of $H_3$ approach $\phi=0$ and $\phi=\pi$. Recalling that $H_3$ is generally described by a pair of geodesics in the AdS$_3$ covering space, we see that one of these geodesics still shrinks to zero coordinate size along the boundary in this limit, though the other part of $H_3$ remains of finite size.


\section{Minimal distance between horizons in the hot limit}\label{appdij}
We now generalize \eqref{dij} to the case of the rotating $(3,0)$ geometry. We focus on the distance $d_{12}$ between $H_1$ and $H_2$ since it is the simplest in our representation of the geometry. Due to the symmetry of the construction, the point on $H_1$ that is closest to $H_2$ sits at the origin of global coordinates. Furthermore, if the point on $H_2$ that is closest to $H_1$ has coordinates $(t_m,r_m,\phi_m)$, then $t_m=0$ by left-right symmetry (see figure \ref{fig:diamondsRot}) and we can set the angular coordinate such that $\phi_m=0$. Recall that any geodesic in AdS$_3$ can be viewed as the intersection of a plane in the embedding space \eqref{pembedd} that passes through the origin with the hyperboloid of AdS$_3$. The idea here is to find the two vectors that span the plane defining $H_2$, then use them to find $r_m$. Using the geodesic distance equation \eqref{Lspacelike}, we can then find $d_{12}$.

Suppose that the left and right corners of the diamond of $H_2$ have coordinates $(-t_0,-\phi_0)$ and $(t_0,\phi_0)$, respectively, at the boundary. Using \eqref{pFixed} and \eqref{H2Diamonds}, it is straightforward to show that
\begin{equation}
t_0 =\tan^{-1} e^{-\alpha}-\tan^{-1} e^{-\tilde{\alpha}}
\end{equation}
\begin{equation}
\phi_0 =\tan^{-1} e^{-\alpha}+\tan^{-1} e^{-\tilde{\alpha}}.
\end{equation}
Then, in embedding space, the vectors $\vec{v}_i=\left(X_i,Y_i,U_i,V_i\right)$ that point from the origin to the points $(-t_0,-\phi_0)$ and $(t_0,\phi_0)$ at the boundary can be found using \eqref{globalAdS3} to be
\be
\vec{v}_L=\left(\cos\phi_0,-\sin\phi_0,\cos t_0,-\sin t_0\right)\quad\text{and}\quad\vec{v}_R=\left(\cos\phi_0,\sin\phi_0,\cos t_0,\sin t_0\right)
\ee
The vector connecting the origin with $(0,r_m,0)$ is parallel to $\vec{v}_L+\vec{v}_R$. From this, it is easy to show that
\be\label{rmH2}
\frac{r_m}{\sqrt{1+r_m^2}}=\frac{\cos\phi_0}{\cos t_0}
\ee
The matrix representation of $(0,r_m,0)$ is
\be
p_m=\left(\begin{array}{cc}
\sqrt{1+r_m^2}+r_m & 0 \\
0 & \sqrt{1+r_m^2}-r_m,
\end{array}\right)
\ee
So, using \eqref{Lspacelike}, the minimal distance between $H_1$ and $H_2$ is the geodesic distance between $p_m$ and the origin and is given by
\be
d_{12}=\cosh^{-1}\left(\frac{\Tr p_m}{2}\right)=\cosh^{-1}\left(\sqrt{1+r_m^2}\right).
\ee
Combining this with \eqref{rmH2} gives
\be\label{dijRot}
d_{12}=\tanh^{-1}\left(\frac{\cos\phi_0}{\cos t_0}\right).
\ee
After some algebra, this can be simplified to
\begin{equation}
\label{explicitd12}
d_{12}= \frac{\alpha+\tilde{\alpha}}{2}.
\end{equation}
which implies that $\alpha,\tilde{\alpha}\geq0$. As a consistency check, note that in the non-rotating case where $\ell_i=\tilde{\ell}_i$, we have
\begin{equation}
\alpha=\tilde{\alpha}\Rightarrow  d_{12}=\alpha,
\end{equation}
which is precisely \eqref{d12alphaEq} as quoted in section \ref{sec2.3}. Other minimal geodesic distances (i.e. $d_{23}$ and $d_{13}$) can be obtained from \eqref{explicitd12} by simple permutations. This completes our generalization of the minimal geodesic distance equation to the rotating case. That the angular domain $D_\phi$ over which $d_{12}$ is exponentially small is also large compared with the AdS length scale in the rotating case follows from the same analysis as in \cite{Marolf:2015vma} through an appropriate choice of the Cauchy slice on which the distance is calculated.

\subsection{The large horizon limit near extremality}
This is the limit where
\begin{align}
 \ell_i\rightarrow\infty\quad\text{and}\quad \tilde{\ell}_i\rightarrow 0 \quad\Leftrightarrow\quad h_i\rightarrow \infty\quad\text{and}\quad T_{H,i}\rightarrow 0.
 \end{align}
From \eqref{explicitd12}, it is easy to see that the above requires
\begin{align}
\alpha\rightarrow 0\quad\text{and}\quad \tilde{\alpha}\rightarrow \infty \quad\Rightarrow\quad  d_{ij}\rightarrow\infty.
\end{align}
This shows that the minimal geodesic distance between the horizons in the extremal limit will diverge. In particular, one can show that the divergence is logarithmic $d_{ij}\sim \log\left(2/\pi T_H\right)+\mathcal{O}\left(T_H^2\right)$. Note however that the hot limit studied in the current  paper instead yields
\begin{align}
 \ell_i\rightarrow\infty\quad\text{and}\quad \tilde{\ell}_i\rightarrow \infty \quad\Leftrightarrow\quad h_i\rightarrow \infty\quad\text{and}\quad T_{H,i}\rightarrow \infty,
 \end{align}
implying that
 \begin{align}
\alpha \rightarrow 0\quad\text{and}\quad \tilde{\alpha} \rightarrow 0 \quad\Rightarrow\quad  d_{ij}\rightarrow 0.
\end{align}
Thus our hot limit implies large horizons, but near extremality large horizons do not imply a hot limit. It also shows that the exponentially small local causal shadow region exists {\it only} in the hot limit where $\alpha$ and $\tilde \alpha$ are both small. 

\bibliographystyle{JHEP}
\cleardoublepage
\renewcommand*{\bibname}{References}
\bibliography{ref}
\end{document}